\documentclass[aps,prd,reprint,preprintnumbers,showpacs,floatfix,nofootinbib,superscript address]{revtex4-2}
\usepackage[utf8]{inputenc}
\usepackage{parskip}
\usepackage{amssymb}
\usepackage{stix}
\usepackage{hhline}
\usepackage{amsmath}
\usepackage{mathtools}
\usepackage[dvipsnames]{xcolor}
\usepackage{xspace}
\usepackage{multirow,tabularx}
\usepackage{siunitx}
\usepackage{multirow}
\usepackage{graphicx}
\usepackage{xstring}
\usepackage{etoolbox}
\usepackage{notoccite}
\usepackage{natbib}
\usepackage{mathrsfs}
\usepackage{lineno}
\usepackage{tensor}
\usepackage{accents}
\usepackage{tikz}
\allowdisplaybreaks

\usepackage{quoting,xparse}
\NewDocumentCommand{\bywhom}{m}{%
	{\nobreak\hfill\penalty50\hskip1em\null\nobreak
		\hfill\mbox{\normalfont(#1)}%
		\parfillskip=0pt \finalhyphendemerits=0 \par}%
}

\NewDocumentEnvironment{pquotation}{m}
{\begin{quoting}[
		indentfirst=true,
		leftmargin=\parindent,
		rightmargin=\parindent]\itshape}
	{\bywhom{#1}\end{quoting}}

\newrobustcmd{\PGT}{%
	{PGT\textsuperscript{q+}}%
}

\newrobustcmd{\LambdaLCDM}{%
	{\Lambda_{\text{LCDM}}}%
}

\newrobustcmd{\Planck}{%
	{M_{\text{Pl}}}%
}

\newrobustcmd{\psic}{%
	{\psi_{\text{C}}}%
}

\newrobustcmd{\LCDM}{%
	{LCDM}%
}

\newrobustcmd{\jfOmegaomega}{%
	{{\hat{\Omega}_{\omega}}}%
}

\newrobustcmd{\jfOmegaM}{%
	{{\hat{\Omega}_{\text{M}}}}%
}
\newrobustcmd{\efOmegaM}{%
	{{\tilde{\Omega}_{\text{M}}}}%
}

\newrobustcmd{\jfF}{%
	{\hat{F}}%
}

\newrobustcmd{\jfB}{%
	{\hat{B}}%
}
\newrobustcmd{\efB}{%
	{\tilde{B}}%
}

\newrobustcmd{\jfrho}{%
	{{\hat{\rho}_{\text{M}}}}%
}
\newrobustcmd{\efrho}{%
	{{\tilde{\rho}_{\text{M}}}}%
}

\newrobustcmd{\jfP}{%
	{\hat{P}_{\text{M}}}%
}
\newrobustcmd{\efP}{%
	{\tilde{P}_{\text{M}}}%
}

\newrobustcmd{\jfH}{%
	{\hat{H}}%
}
\newrobustcmd{\efH}{%
	{\tilde{H}}%
}

\newrobustcmd{\jfx}{%
	{\hat{x}}%
}
\newrobustcmd{\efx}{%
	{\tilde{x}}%
}

\newrobustcmd{\jfy}{%
	{\hat{y}}%
}
\newrobustcmd{\efy}{%
	{\tilde{y}}%
}

\newrobustcmd{\jfz}{%
	{\hat{z}}%
}
\newrobustcmd{\efz}{%
	{\tilde{z}}%
}

\newrobustcmd{\jfL}[1]{%
	{\hat{L}_{\text{#1}}}%
}
\newrobustcmd{\efL}[1]{%
	{\tilde{L}_{\text{#1}}}%
}

\newrobustcmd{\jfX}[1]{%
	\tensor{\hat{X}}{^{#1}}%
}
\newrobustcmd{\efX}[1]{%
	\tensor{\tilde{X}}{^{#1}}%
}

\newrobustcmd{\jfphi}[1]{%
	\tensor{\hat \phi}{#1}%
}
\newrobustcmd{\jfpsi}[1]{%
	\tensor{\hat \psi}{#1}%
}

\newrobustcmd{\jfR}[1]{%
	\tensor{\hat R}{#1}%
}

\newrobustcmd{\jfg}[1]{%
	\tensor{\hat g}{#1}%
}
\newrobustcmd{\efg}[1]{%
	\tensor{\tilde g}{#1}%
}

\newrobustcmd{\jfOmega}{%
	{\hat \Omega}%
}
\newrobustcmd{\efOmega}{%
	{\tilde \Omega}%
}

\newrobustcmd{\jfD}[1]{%
	\tensor{\hat{\nabla}}{#1}%
}

\newrobustcmd{\jfomega}[1]{%
	\tensor{\hat \omega}{#1}%
}
\newrobustcmd{\efpsi}[1]{%
	\tensor{\tilde \omega}{#1}%
}

\newrobustcmd{\efR}[1]{%
	\tensor{\tilde R}{#1}%
}

\newrobustcmd{\jfU}[1]{%
	\tensor{\hat U}{#1}%
}
\newrobustcmd{\efU}[1]{%
	\tensor{\tilde U}{#1}%
}

\newrobustcmd{\gef}[1]{%
	\tensor{\tilde g}{#1}%
}

\newrobustcmd{\efD}[1]{%
	\tensor{\tilde{\nabla}}{#1}%
}

\newrobustcmd{\rD}[1]{%
	\tensor{\mathring{\nabla}}{#1}%
}

\newrobustcmd{\rcD}[1]{%
	\tensor{\nabla}{#1}%
}

\newrobustcmd{\rR}[1]{%
	\tensor{\mathring{R}}{#1}%
}

\newrobustcmd{\rcR}[1]{%
	\tensor{R}{#1}%
}

\newrobustcmd{\T}[2][placeholder]{%
	\IfEqCase{#1}{%
	{placeholder}{\tensor{T}{#2}}%
	{1}{\tensor[^{(1)}]{T}{#2}}%
	{2}{\tensor[^{(2)}]{T}{#2}}%
	{3}{\tensor[^{(3)}]{T}{#2}}%
	}%
	[\packageError{cosmicclass}{Symbol #1 is not an irreducible part!}{}]%
}

\newrobustcmd{\TLambda}[2][placeholder]{%
	\IfEqCase{#1}{%
	{placeholder}{\tensor{\lambda}{#2}}%
	{1}{\tensor[^{(1)}]{\lambda}{#2}}%
	{2}{\tensor[^{(2)}]{\lambda}{#2}}%
	{3}{\tensor[^{(3)}]{\lambda}{#2}}%
	}%
	[\packageError{cosmicclass}{Symbol #1 is not an irreducible part!}{}]%
}

\def\aap{\rm{A\&A}}

\def\prd{\rm{Phys.~Rev.~D}}

\def\jmp{\rm{J.~Math.~Phys.}}

\def\ijmpd{\rm{Int.~J.~Mod.~Phys.~D}}

\def\pletta{\rm{Phys.~Lett.~A}}

\def\progtp{\rm{Prog.~Theor.~Phys.}}

\usepackage{hyperref}
\hypersetup{%
	colorlinks = true,%
	linkcolor = Blue,%
	citecolor = Blue,%
	filecolor = Blue,%
	urlcolor = Blue%
}%
\usepackage[capitalize]{cleveref} 

\begin{document}
	
\title{The effective inflationary potential of constant-torsion emergent gravity}

\author{C. Rew}
\email{srcx26@durham.ac.uk}
\affiliation{Department of Physics, Durham University, Lower Mountjoy, South Road, Durham DH1 3LE, UK}
\author{W.E.V. Barker}
\email{wb263@cam.ac.uk}
\affiliation{Astrophysics Group, Cavendish Laboratory, JJ Thomson Avenue, Cambridge CB3 0HE, UK}
\affiliation{Kavli Institute for Cosmology, Madingley Road, Cambridge CB3 0HA, UK}


\begin{abstract}
	Constant-torsion emergent gravity (CTEG) has a Lagrangian quadratic in curvature and torsion, but without any Einstein--Hilbert term. CTEG is motivated by a unitary, power-counting renormalisable particle spectrum. The timelike axial torsion adopts a vacuum expectation value, and the Friedmann cosmology emerges dynamically on this torsion condensate. We show that this mechanism -- and the whole background cosmology of CTEG -- may be understood through the effective potential of a canonical single scalar field model. The effective potential allows for hilltop inflation in the early Universe. In the late Universe, the Hubble friction overdamps the final quadratic approach to the effective minimum at the condensate, where the value of the potential becomes the cosmological constant. 
	We do not consider particle production through spin-torsion coupling, or running of Lagrangian parameters. The model must be completed if reheating and a separation of inflationary and dark energy scales are to be understood. It is suggested that the divergence of the potential at large values of the scalar is inconsistent with the linearised propagator analysis of CTEG around zero-torsion Minkowski spacetime. This background may therefore be a strongly coupled surface in CTEG.
\end{abstract}


\maketitle

\section{Introduction}\label{sec: Intro}

The early and late Universe are both characterised by accelerated expansion with a nearly constant Hubble number. These regimes may be realised in general relativity (GR) by means of a slowly-rolling inflaton field and, in the full cosmic concordance (\LCDM{}) model~\cite{planck2014,2020A&A...641A..10P}, with a small positive external cosmological constant $\LambdaLCDM{} \approx 2.846 \times 10^{-122} \Planck^2$~\cite{2014A&A...571A..16P,2018arXiv180706209P}. It is then interesting to consider whether \emph{modified} gravity models can furnish such driving mechanisms from within the gravitational sector itself.

\subsubsection{Poincar\'e gauge theory}

GR may be naturally extended by augmenting the
Levi--Civita connection $\tensor{\Gamma}{^\mu_{\nu\sigma}}=\tensor*{C}{^\mu_{\nu\sigma}}\equiv\tfrac{1}{2}\tensor{g}{^{\mu\rho}}\tensor{\partial}{_\nu}\tensor{g}{_{\rho\sigma}}+...$ with an antisymmetric \emph{contortion} correction. Contortion directly encodes spacetime \emph{torsion} $\tensor{\mathcal{T}}{^\mu_{\nu\sigma}}\equiv2\tensor{\Gamma}{^\mu_{[\nu\sigma]}}$, whose presence may eventually be indicated
by a breaking of the equivalence principle within the fermionic sector~\cite{2014IJMPD..2342004P},
depending on the detailed (and entirely unknown) nature of the matter coupling\footnote{Note that some authors have also suggested that non-Riemannian spacetime geometries may, for specific gravity models, be distinguished by the geodesic or autoparallel character of particle trajectories~\cite{Acedo:2015fcf}.}.
Torsion theories can be motivated within the traditional GR framework by
promoting the spin connection to an independent gauge field of the Lorentz 
group. This procedure generates the celebrated Poincar\'e gauge theory of gravity (PGT)~\cite{RevModPhys.36.463,kibble,PhysRev.101.1597}, in which the eighteen independent non-gauge components of the spin connection may propagate alongside the graviton as scalar $0^\pm$, vector $1^\pm$ or tensor $2^\pm$ particles~\cite{10.1143/PTP.64.866,10.1143/PTP.64.1435,10.1143/PTP.64.2222}, depending on the specific balance of quadratic curvature ($\mathcal{R}^2$) invariants present in the low energy expansion. The external masses of these particles are contingent on the Einstein--Hilbert ($\mathcal{R}$) and quadratic torsion ($\mathcal{T}^2$) coupling constants; the low energy theory up to \emph{even} parity \emph{quadratic} invariants of curvature and torsion is denoted \PGT{}.

Stringent constraints on the \PGT{} establish that only the scalar $0^\pm$ modes may propagate without exciting ghosts in the fully nonlinear theory~\cite{2002IJMPD..11..747Y,1999IJMPD...8..459Y}. However, these constraints are only known to apply to theories which \emph{modify} GR, for instance, as low energy effective theories $L\sim\mathcal{R}+\mathcal{R}^2+\mathcal{T}^2+L_{\text{M}}$ which modify the Einstein--Cartan (EC) model $L\sim\mathcal{R}+L_{\text{M}}$, where the matter Lagrangian is $L_{\text{M}}$. Recent work has shown that there is a discrete collection of \PGT{} actions whose (linearised) particle spectra are free from ghosts and tachyons~\cite{Lin1}, and which actually appear renormalisable by a power counting~\cite{Lin2}. It is hard to see how renormalisable, unitary gravity models can be consistently obtained as `extra-particle' modifications of GR, and indeed these external actions are \emph{purely quadratic in curvature and torsion}, entirely lacking an Einstein--Hilbert term $L\sim\mathcal{R}^2+\mathcal{T}^2+L_{\text{M}}$. It is remarkable that from this motivation in the UV, the nonlinear, classical phenomenology of these models, when coupled minimally to matter, can still admit a wholly viable background cosmology~\cite{chapter2,Barker_2020,mythesis}.

In the general \PGT{}, it follows from the isotropy of the Friedmann--Lema\^itre--Robertson--Walker (FLRW) model that only the $0^\pm$ modes do not vanish at homogeneous scales~\cite{TSAMPARLIS197927}. For \PGT{} theories which \emph{modify} GR, these scalars may play a natural role in inflationary and dynamical dark energy models, by immediate analogy to scalar-tensor theories of gravity. Indeed, those `permitted' low-energy \PGT{}s in which the massive $0^\pm$ modes alone are propagating, are dynamically equivalent to scalar-tensor gravity without torsion~\cite{10.1143/PTP.64.866,blagojevic2002gravitation,mythesis}.

\subsubsection{The CTEG model}\label{TheCTEGModel}

The new unitary, putatively renormalisable $L\sim\mathcal{R}^2+\mathcal{T}^2+L_{\text{M}}$ \PGT{} is most fully developed in the \emph{constant-torsion emergent gravity} (CTEG) model~\cite{chapter2,Barker_2020,mythesis}. The Lagrangian for this theory is
\begin{widetext}
	\begin{equation}
		\begin{aligned}
			L_\text{CTEG}&=
			-\frac{4\Planck^2}{9}
			\tensor{\mathcal{  T}}{_\mu}\tensor{\mathcal{  T}}{^\mu}
			-\frac{\mu}{6}\Big[
			\lambda\tensor{\mathcal{  T}}{_{\mu\nu\sigma}}
			\big(
			\tensor{\mathcal{  T}}{^{\mu\nu\sigma}}
			-2\tensor{\mathcal{  T}}{^{\nu\mu\sigma}}
			\big)
			+\tensor{\mathcal{  R}}{_{\mu\nu}}
			\big(
			\tensor{\mathcal{  R}}{^{[\mu\nu]}}
			-12\tensor{\mathcal{  R}}{^{\mu\nu}}
			\big)
			-2\tensor{\mathcal{  R}}{_{\mu\nu\sigma\rho}}
			\big(
			\tensor{\mathcal{  R}}{^{\mu\nu\sigma\rho}}
			-4\tensor{\mathcal{  R}}{^{\mu\sigma\nu\rho}}
			-5\tensor{\mathcal{  R}}{^{\sigma\rho\mu\nu}}
			\big)
			\Big]
			\\
			&\ \ \ \ 
			+2\nu\tensor{\mathcal{  R}}{_{[\mu\nu]}}\tensor{\mathcal{  R}}{^{[\mu\nu]}}
			-\Planck^2\Lambda+L_{\text{M}},
			\label{CTEG}
		\end{aligned}
	\end{equation}
\end{widetext}
in which $\tensor{\mathcal{  R}}{_{\mu\nu\sigma\rho}}$ is the Riemann--Cartan (i.e. torsionful curvature, with its reduced symmetries) tensor\footnote{Our convention for the torsion-free Riemann tensor will be $\tensor{R}{^\mu_{\nu\sigma\rho}}\equiv 2\tensor{\partial}{_{[\sigma|}}\tensor*{C}{^\mu_{|\rho]\nu}}+\dots$, with Ricci scalar $R\equiv\tensor{R}{^{\mu\nu}_{\mu\nu}}$, and for the Riemann--Cartan tensor $\tensor{\mathcal{R}}{^\mu_{\nu\sigma\rho}}\equiv 2\tensor{\partial}{_{[\sigma|}}\tensor{\Gamma}{^\mu_{|\rho]\nu}}+\dots$, with contractions $\tensor{\mathcal{R}}{_{\mu\nu}}\equiv\tensor{\mathcal{R}}{^{\sigma}_{\mu\sigma\nu}}$ and $\mathcal{R}\equiv\tensor{\mathcal{R}}{^{\mu}_{\mu}}$.}, and $\tensor{\mathcal{  T}}{_{\mu\nu\sigma}}$ is the torsion tensor, with $\tensor{\mathcal{  T}}{_{\nu}}\equiv\tensor{\mathcal{  T}}{^\mu_{\nu\mu}}$. The modified gravitational sector is assumed to be minimally coupled to the standard model through the conventional matter Lagrangian $L_{\text{M}}$.

Apart from the Planck mass $\Planck\equiv\SI[separate-uncertainty=true]{2.435320(28)e27}{\electronvolt}$, four parameters appear in the theory~\eqref{CTEG}. These are the dimensionless $\mu$ and $\nu$, and the dimensionful `cosmological constants' $\lambda$ and $\Lambda$. The unitarity of the free gravity theory~\cite{Lin2}, in which the externally imposed $\Lambda$ and $L_{\text{M}}$ are neglected, requires
\begin{equation}\label{UnitarityConditions}
	\lambda\geq 0, \quad \mu<0, \quad (\nu+2\mu)(\nu-\mu)>0,
\end{equation}
and is evaluated on a flat Minkowski background where both curvature and torsion are vanishing.
\subsubsection{Torsion condensation and the CS}
Exact solutions have not, to date, constrained the dimensionless $\mu$ and $\nu$. However the late Universe is known to transition to an asymptotically de Sitter expansion, with constant Hubble number $H\to\sqrt{(\Lambda+\lambda)/3}$. Thus, the accelerated expansion is accounted for by an `external' or `geometric' cosmological constant $\Lambda$, such as is usually included in the Einstein--Hilbert Lagrangian, summed with the torsion-torsion coupling $\lambda$. Contrast this with the case of GR
\begin{equation}\label{GR}
	L_{\text{GR}}=-\frac{\Planck^2}{2}R-\Planck^2\Lambda+L_{\text{M}},
\end{equation}
in which only the external cosmological constant, included via Lovelock's theorem, is felt in the late Universe $H\to\sqrt{\Lambda/3}$.

In fact the cosmology of the theory~\eqref{CTEG} is one of its more intriguing features. Despite lacking an Einstein--Hilbert term as in~\eqref{GR}, precisely the Friedmann equations of GR emerge when the scalar and pseudoscalar parts of the torsion evolve according to
\begin{equation}\label{TorsionCondensate}
	\phi=\phi(t), \quad \psi=\psi_{\text{C}}\equiv\frac{\Planck}{\sqrt{-3\mu}},
\end{equation}
where the scalars are respectively the $0^+$ and $0^-$ modes defined by the unit vector $\tensor{n}{^\mu}\tensor{n}{_\mu}\equiv 1$ in the direction of cosmic time $t$, according to the well-known homogeneous, isotropic torsion ansatz~\cite{1979PhLA...75...27T}
\begin{equation}\label{TsamparlisFormula}
	\tensor{\mathcal{T}}{^\mu_{\nu\sigma}}=(\phi+2H)\tensor*{\delta}{^\mu_{[\sigma}}\tensor{n}{_{\nu]}}
	+\psi\tensor{n}{_{\rho}}\tensor{\epsilon}{^{\rho\mu}_{\nu\sigma}},
\end{equation}
where $H$ is the Hubble number. In this work we assume a flat FLRW background, with the line element
\begin{equation}\label{metric}
	\mathrm{d}s^2 = \mathrm{d}t^2 - a(t)^2 \mathrm{d} \mathbf{x}^2,
\end{equation}
with $a$ being the cosmological scale factor, such that the Hubble number is $H \equiv \Dot{a}/a$. 

When the cosmic fluid has a constant equation of state (e.o.s) parameter $w\equiv P/\rho$, the first equality in~\eqref{TorsionCondensate} may be further resolved to $\phi\propto H$, where the dimensionless constant of proportionality is determined by $w$. We recall $w=1/3$ and $H=1/2t$ while the species of $L_{\text{M}}$ are radiative, after which time $w=0$ and $H=2/3t$ for a cold dust (baryons and cold dark matter). The late Universe will witness a transition to $w=-1$ as $L_{\text{M}}$ dilutes and becomes less important, and eventually $H=\sqrt{(\Lambda+\lambda)/3}$. While $\phi$ continues to evolve through the cosmic history, $\psi=\psi_{\text{C}}$ remains fixed and maintains the emergent Friedmann equations: we refer to this as the \emph{torsion condensate} or \emph{correspondence solution} (CS).

In~\eqref{TorsionCondensate} the coupling $\nu$ makes no appearance, whilst $\mu$ is absorbed into the definition of $\psi_{\text{C}}$. 
The Friedmann equations of GR do not, of course, refer to $\psi_{\text{C}}$, but only to $H$ and the densities and pressures imposed by $\Lambda+\lambda$ (note the CTEG combination) and $L_{\text{M}}$. Accordingly, we find that analysis of the expansion history is not sensitive to the value of $\mu$, and in fact this will turn out to be true even away from the condensate so long as the dynamical equations are expressed in terms of the natural ratio $\psi/\psi_{\text{C}}$. The couplings $\mu$ and $\nu$ in~\eqref{CTEG} thus drop out of the picture entirely, and for the remainder of this work the CTEG will be parameterised \emph{entirely} by the two cosmological constant parameters $\lambda$ and $\Lambda$.

\subsubsection{Open problems}
There are strong indications that the CS is stable against perturbations at the background level~\cite{chapter2,mythesis}. Deviations in $\psi$ decay most slowly towards the condensate $\psi_{\text{C}}$ during the radiation-dominated epoch, allowing~\eqref{CTEG} to modify the \LCDM{} thermal history according to a single parameter: the initial value of $\psi$ at the end of reheating. If initially $\psi\lesssim\psi_{\text{C}}$, the net effect is equal to that of so-called `dark radiation' models, which boost the early-Universe expansion rate. Such models have been proposed in order to alleviate the Hubble tension.

It is not yet clear \emph{how} the CS is formed in the first place. The CTEG particle spectrum was initially obtained~\cite{Lin2} on the non-expanding Minkowski background, with vanishing torsion $\psi=\phi=H=0$. There was moreover no external cosmological constant or matter source present, so $\Lambda=L_{\text{M}}=0$, though the spectra of CTEG both with and without the coupling $\lambda$ were computed: the spectra differ by the massive $0^-$ mode $\psi$, which becomes non-propagating if $\lambda=0$ and tachyonic if $\lambda<0$, consistent with~\eqref{UnitarityConditions}. Both these versions of free CTEG contain two massless polarizations of unknown spin and parity, which we consider to be the \emph{gravitons}. It is currently understood that the CS is an inherently \emph{nonlinear} feature of the dynamics, which is not captured by the perturbative propagator analysis: this is evidenced by the fact that the CS can form as a vacuum expectation value $\psi=\psi_{\text{C}}$ of the $0^-$ mode with both $\lambda>0$ and $\lambda=0$ versions of CTEG, i.e. \emph{regardless of whether $\psi$ is even supposed to be propagating}. This picture is disconcertingly familiar from the theory of strongly coupled surfaces~\cite{BeltranJimenez:2020lee,smooth}. Indeed, a preliminary Hamiltonian analysis of the CTEG in~\cite{mythesis} indicated the presence of strongly coupled modes around the original $\psi=\phi=H=\Lambda=L_{\text{M}}=0$ background, though these results were not fully conclusive.

Even if the $\psi=\phi=H=\Lambda=L_{\text{M}}=0$ background is strongly coupled, CTEG may yet remain viable if the CS itself is (i) not strongly coupled and (ii) also furnished with an equally attractive particle spectrum.
The focus then turns to inhomogeneous, anisotropic cosmological perturbations around~\eqref{TorsionCondensate} --- including perturbations of the remaining sixteen polarisation components of the spin connection --- and the propagator structure in that environment. This is an area of current investigation~\cite{chapter4,mythesis}, but there is already some suggestion that the Newtonian limit of local overdensities can be recovered~\cite{mythesis}. The need to avoid strong coupling of the torsion condensate brings us back to the question of \emph{how} the CS may be formed, i.e. whether it can be reached dynamically as a non-singular surface in the phase space. If it can be formed, and if that process of formation occurs early enough in the history of the Universe, then it would be quite advantageous if the `condensation' process happened to produce 50-60 e-folds of inflation as a by-product. 

\subsubsection{Results of this work}

This work relates the CTEG Lagrangian~\eqref{CTEG} with the `reduced' model, to be expressed finally in~\cref{SummaryOfModel,TransformationToModel}. This model fully preserves the background cosmological dynamics entailed by~\cref{TsamparlisFormula,metric}.

We will start with a recapitulation of the work conducted in \cite{Barker_2020}, with particular focus on the bi-scalar-tensor analogue Lagrangian for the general \PGT{} and specific CTEG theories. Then we will show the equivalence of the $\Lambda=0$ scalar-tensor analogue with a non-minimally coupled single scalar field Lagrangian in \cref{sec: reduction to single SF theory}. This will be followed by a rigorous dynamical systems analysis of the new Lagrangian in \cref{sec: dyn sys in JF}; starting with the construction of the phase space in \cref{sec:constructing phase space in JF}. Using this we will show the stability of the CS in \cref{sec: stability in JF}, thus concluding the analysis of the theory in the Jordan frame.

Following the analysis of the theory in the Jordan frame, we will move the focus to the Einstein frame formulation of the theory, by performing a conformal transformation in \cref{sec:EF}. We will repeat the dynamical systems analysis in the Einstein frame, and show that both of the theories have stable points at the CS.

Finally, we will look at the effect of adding an external cosmological constant $\Lambda$ in the original Jordan frame scalar-tensor analogue. In this section we will give some discussion on the range of values the external cosmological constant $\Lambda$ can take from a dynamical systems point of view, as well as a look at the effects of $\Lambda$ on the potential for inflation. We will show that there are values that the two cosmological constants of the theory ($\Lambda$ and $\lambda$) can take such that there is an inflationary hilltop regime which produces at least 50 e-folds of inflation in \cref{sec:inflation}. Whilst the shape of the potential is suitable, the scale of inflation is insufficient when the late-time Hubble number is imposed. The resulting model is incomplete without an understanding of the particle theory, but serves as a qualitative test of inflation in CTEG. Conclusions follow in~\cref{Conclusions}, and we reiterate that this final section contains~\cref{SummaryOfModel,TransformationToModel} which encode our main result.

We will use the metric signature $(+,-,-,-)$ in the rest of the work. A list of nonstandard acronyms is provided in~\cref{NonstandardAcronyms}.

\begin{table}[t!]
	\caption{\label{NonstandardAcronyms} Nonstandard acronyms used in this work.}
	\begin{center}
		\begin{tabularx}{\linewidth}{c|l}
			\hline\hline
			\PGT{} & Parity-preserving, quadratic Poincar\'e gauge theory \\
			CTEG & Constant-torsion emergent gravity, defined in~\eqref{CTEG} \\
			CS & Correspondence solution, i.e. torsion condensate $\psi=\psi_{\text{C}}$ \\
			e.o.s & Equation of state \\
			e.o.m & Equation of motion \\
			\hline\hline
		\end{tabularx}
	\end{center}
\end{table}

\section{The bi-scalar-tensor theory}

The restriction of attention to the scalar $0^\pm$ modes~\eqref{TsamparlisFormula} when considering the cosmological background invites the construction of a torsion-free, scalar-tensor theory which can fully replicate the cosmological background of general \PGT{}. This scalar-tensor analogue theory was identified in~\cite{Barker_2020}: it has in general an Einstein--Hilbert term, non-minimally coupled to a pair of scalar fields $\phi$ and $\psi$ which emulate the dynamics of~\eqref{TsamparlisFormula} at the background level, and whose kinetic terms may be non-canonical. Curiously, the Einstein--Hilbert term in the analogue does not directly translate to the Einstein--Hilbert term in the \PGT{}. In the limiting cases, the zero-curvature ($\mathcal{R}=0$) \emph{teleparallel equivalent of GR} (TEGR) has an analogue $L_{\text{TEGR}}=-\frac{1}{2}\Planck^2R+L_{\text{M}}$ of pure GR without any scalars, whilst the conservative Einstein--Cartan (EC) model has an analogue of a pure, massive \emph{Cuscuton} field\footnote{See~\cite{afshordi2007cuscuton} for an introduction to the Cuscuton.} $L_{\text{EC}}=-{\Planck}^2\sqrt{|2X^{\phi\phi}|}+\tfrac{3}{4}{\Planck}^2\phi^2+L_{\text{M}}$, with $X^{\phi\phi}\equiv\frac{1}{2}\tensor{g}{^{\mu\nu}}\tensor{\partial}{_\mu}\phi\tensor{\partial}{_{\nu}}\phi$. Even without any Ricci scalar, the quadratic Cuscuton still supports the Friedmann equations at the background level\footnote{To see this, substitute the $\phi$-equation into the very simple $\tensor{g}{^{\mu\nu}}$-equation to recover the Friedmann constraint equation.}.

The scalar-tensor analogue action from \cite{Barker_2020} corresponding to the CTEG in~\eqref{CTEG} is more complex. We will begin with the case $\Lambda=0$, and re-introduce the external cosmological constant only from~\cref{ReintroduceLambda} onwards. The scalar-tensor analogue may be brought to a minimally coupled frame via a conformal transformation $\tensor{g}{_{\mu\nu}}\equiv\jfOmega{}(\psi)^2\jfg{_{\mu\nu}}$, followed by field redefinitions $\phi\equiv\phi\left(\jfphi{},\jfpsi{}\right)$ and $\psi\equiv\psi\left(\jfpsi{}\right)$, where
\begin{subequations}
	\begin{align}
		\jfOmega(\psi)^2&\equiv3\left(4-\frac{\psi^2}{{\psi_{\text{C}}}^2}\right)^{-1},\label{JFOmega}\\
		\phi(\jfphi{},\jfpsi{})&\equiv\jfphi{}\sqrt{\frac{8\alpha}{3}\left[3\cosh\left(\sqrt{\frac{2}{3}}\frac{\jfpsi{}}{\Planck}\right)-5\right]}\nonumber\\
		&\quad\quad\quad\quad\quad\quad\quad\quad\quad\quad \times\mathrm{sech}\left(\frac{\jfpsi{}}{\sqrt{6}\Planck}\right),
		\\
		\psi(\jfpsi{})&\equiv 2\psi_{\text{C}}\tanh\left(\frac{\jfpsi{}}{\sqrt{6}\Planck}\right),\label{PsiOfXi}
	\end{align}
\end{subequations}
where we keep track of the sign
\begin{equation}\label{AlphaDefinition}
	\alpha \equiv \mathrm{sgn} \left[3\cosh \left(\sqrt{\frac{2}{3}} \frac{\jfpsi{}}{\Planck}\right)-5 \right].
\end{equation}
It is important to note that we will refer to the frame $\jfg{_{\mu\nu}}$ as the \emph{Jordan} frame, because it is only minimally coupled before the field $\jfphi{}$ has been integrated out in~\cref{sec: reduction to single SF theory}. A non-minimal coupling between $\jfomega{}\left(\jfpsi{}\right)$ and $\jfR{}$ will then be induced. In~\cref{sec:EF} a \emph{second} conformal transformation will be introduced to decouple the scalar, and only this final frame --- two conformal transformations removed from the physical frame $\tensor{g}{_{\mu\nu}}$ of~\cref{CTEG} --- will be referred to as the \emph{Einstein} frame.
Note from~\cref{JFOmega,PsiOfXi}, the conformal transformation only admits the range $\psi\in\left(-2\psic,2\psic\right)$, which fills the whole range $\jfpsi{}\in\left(-\infty,\infty\right)$. We do not need to consider values of the axial torsion \emph{significantly} above the condensate level in this work, and within the physical range we may conclude
\begin{equation}\label{AlphaDefinition}
	\alpha=
	\begin{cases}
		-1, & |\psi|<\psi_{\text{C}},\\
		1, & \psi_{\text{C}}<|\psi|.
	\end{cases}
\end{equation}
Using~\crefrange{JFOmega}{AlphaDefinition}, the bi-scalar-tensor equivalent of~\eqref{CTEG} in the new frame becomes
\begin{subequations}
	\begin{align}
		\jfL{CTEG} &= -\frac{1}{2} \Planck ^2 \jfR{} +\jfX{\jfpsi{}\jfpsi{}} + \beta\Planck ^2 \jfomega{}(\jfpsi{})^3 \sqrt{\left| \jfX{\jfphi{} \jfphi{}} \right|}\nonumber\\
		&\ \ \ -\jfU{}(\jfpsi{})+  \frac{3}{4} \alpha\Planck ^2 \jfomega{}(\jfpsi{})^4 \jfphi{}^2 +\jfL{M},
		\label{cuscuton lag}\\
		\jfU{}(\jfpsi{}) &\equiv \lambda \Planck ^2 \left(1+\alpha \frac{\jfomega{}(\jfpsi{})^2}{2}\right) \left(1+\alpha \frac{\jfomega{}(\jfpsi{})^2}{8} \right),\\
		\jfomega{}(\jfpsi{}) &\equiv \alpha \sqrt{\left| 3 \cosh \left(\sqrt{\frac{2}{3}} \frac{\jfpsi{}}{\Planck} \right)-5 \right|},
		\label{omega eq}
	\end{align}
\end{subequations}
where the kinetic terms are $\jfX{\jfphi{} \jfphi{}} \equiv \frac{1}{2} \jfg{^{\mu \nu}} \partial_{\nu} \jfphi{} \partial_{\mu} \jfphi{}$ , $\jfX{\jfpsi{} \jfpsi{}} \equiv \frac{1}{2} \jfg{^{\mu \nu}} \partial_{\nu} \jfpsi{} \partial_{\mu} \jfpsi{}$, with $\beta\equiv\mathrm{sgn}\left(\jfX{\jfphi{} \jfphi{}}\right)$. The $\jfpsi{}$ field has the hallmarks of a canonical scalar field, whereas the $\jfphi{}$ field is a quadratic Cuscuton field. The new treatment of this work is to not only recalculate the dynamics, but also to take into account values of the torsion above and below the correspondence solution; this is achieved by  the appearance of the $\alpha$ and $\beta$ terms in \crefrange{cuscuton lag}{omega eq}. The $\alpha$ and $\beta$ terms track, respectively, the sign of the term inside the modulus in \cref{omega eq} and the sign of the Cuscuton velocity $\Dot{\jfphi{}}$.

The presence of the Cuscuton field in the Lagrangian \cref{cuscuton lag} is a point of concern, as square roots of kinetic energy-like terms are physically questionable and difficult to motivate. The Cuscuton field, as outlined in \cite{afshordi2007causal,afshordi2007cuscuton}, is a non-dynamical field with infinite speed of sound.  The simplicity of scalar field models of dark energy is very appealing, as is reducing \cref{cuscuton lag} down to a single scalar field model, without the phenomenologicaly interesting, but nonetheless irksome, Cuscuton. 

Emphasis is placed on the fact that the model is only valid at the background level, and at this point it is useful to discuss the idea of the correspondence solution (CS). This is an attractive feature of the theory as it is the point at which the scalar-tensor analogue exactly matches the Lagrangian of GR with a cosmological constant. The stability of the CS, which is shown in this work, is an important feature of the theory, as at the background level GR is a good model of the Universe at the current accelerating dark energy dominated epoch.

The physical motivation for reformulating this theory is from inspection of the degrees of freedom (d.o.f) for the Lagrangian in \cref{cuscuton lag}. From \cite{afshordi2007causal,afshordi2007cuscuton} the Cuscuton field $\jfphi{}$ has no propagating d.o.f, rather is acts as a constraint field and has the unusual property that the kinetic term in the Lagrangian does not contribute to the energy density of the field. Therefore, for the interests of being able to analyse the dynamics of the system using the powerful dynamical systems framework, it is convenient to reduce this two field theory to a single scalar field model, with the corresponding single d.o.f.

Variation of \cref{cuscuton lag} with repsect to $\jfpsi{}$, $\jfphi{}$ and $\jfg{^{\mu \nu}}$ gives the following field equations for the bi-scalar-tensor system (where an overdot represents derivative w.r.t cosmic time in the new frame and $'$ denotes a derivative w.r.t the $\jfpsi{}$ scalar field)

\begin{subequations}	
	\begin{align}
		3&\jfH{}^2 \left(1+\alpha \frac{\jfomega{}^2}{2}\right)=\frac{1}{\Planck^2} \left(\frac{1}{2} \Dot{\jfpsi{}}^2 +\jfU{}(\jfpsi{})\right)\nonumber\\
		&+\alpha\left(-3 \jfH{} \jfomega{} \jfomega{}^{\prime}-\frac{3}{2} \alpha \Dot{\jfpsi{}}^2 \jfomega{}^{\prime 2}  \right) , 
		\label{fried 1 cusc}
		\\
		\Big(&2\Dot{\jfH{}} +3\jfH{}^2\Big)\left(1+\alpha \frac{\jfomega{}^2}{2}\right) = -\frac{1}{\Planck^2} \left(\frac{1}{2} \Dot{\jfpsi{}}^2 -\jfU{}(\jfpsi{}) \right) \nonumber\\ 
		+&\alpha\left(-2  \jfH{}\jfomega{}
		\jfomega{}^{\prime} \Dot{\jfpsi{}}+\frac{1}{2} \Dot{\jfpsi{}}^2 \jfomega{}^{\prime 2} -\jfomega{} \jfomega{}^{\prime} \Ddot{\jfpsi{}} -\jfomega{} \jfomega{}^{\prime \prime} \Dot{\jfpsi{}} ^2\right) , 
		\label{fried 2 cusc}
		\\
		\Ddot{\jfpsi{}}& +3\jfH{} \Dot{\jfpsi{}} +\jfU{} ^{\prime} (\jfpsi{}) = \alpha \Planck^2 \Big( \Big. 6\jfH{}^2 \jfomega{} \jfomega{}^{\prime} +3 \Dot{\jfH{}} \jfomega{} \jfomega{}^{\prime} \nonumber\\
		&+ 9\jfH{} \Dot{\jfpsi{}} \jfomega{}^{\prime 2} +3 \jfomega{}^{\prime 2}  \Ddot{\jfpsi{}} + 3 \Dot{\jfpsi{}}^2 \jfomega{}^{\prime} \jfomega{}^{\prime \prime} \Big. \Big).
		\label{kg cusc}
	\end{align}
	These equations are found by eliminating $\jfphi{}$ from the system by substituting in the equation of motion, found by variations of \cref{cuscuton lag} w.r.t $\jfphi{}$
	\begin{align}
		\jfomega{}^2 \left(\sqrt{2} \beta \jfomega{}' \Dot{\jfpsi{}} +\sqrt{2} \beta \jfomega{} \jfH{} - \alpha \jfomega{}^2 \jfphi{}\right) = 0.
	\end{align}
\end{subequations}
Interestingly, upon substitution for the Cuscuton into the field equations, the prefactor of the Cuscuton kinetic term $\beta$ only appears as $\beta^2=1$, so the treatment by most of the literature with regards to the Cuscuton kinetic term's sign not being a necessary part of one's treatment of the system is at least justified for our present case.  

\section{Reduction from a bi-scalar-tensor theory to single scalar field model}\label{sec: reduction to single SF theory}

The starting point for removing the $\jfphi{}$ field follows \cite{afshordi2007causal} where the authors mention that the Cuscuton is a minimal modification to GR at the background level, and showed the Cuscuton action was equivalent to a renormalisation of the Planck mass. This motivates an attempt to remove the explicit Cuscuton field through the single field Lagrangian
\begin{align}
	\jfL{CTEG} &= -\frac{1}{2} \Planck ^2 \jfF{}(\jfpsi{}) \jfR{} +\jfX{\jfpsi{}\jfpsi{}}-\jfU{}(\jfpsi{})\nonumber\\
	&\ \ \ - \alpha \frac{3}{2} \Planck ^2\jfg{^{\mu\nu}} \partial_{\nu} \jfomega{}(\jfpsi{}) \partial_{\mu} \jfomega{}(\jfpsi{}) +\jfL{M},
	\label{first equi lag}
\end{align}
where $\jfF{}(\jfpsi{})$ is at this point some function of the $\jfpsi{}$ field; this will be defined explicity upon the field redefinition \cref{xi to omega redef}. The field equations resulting from \cref{first equi lag} are
\begin{subequations}
	\begin{align}
		3&\jfH{}^2 \jfF{} = \frac{1}{\Planck^2} \left(\frac{1}{2} \Dot{\jfpsi{}}^2 +\jfU{}(\jfpsi{}) \right) + \nonumber\\ 
		&\quad \alpha \left( -3\jfH{}\jfF{}^{\prime} \Dot{\jfpsi{}} -\frac{3}{2} \Dot{\jfpsi{}}^2 {\jfomega{}^{\prime}} ^2 \right) +\jfrho{}, 
		\label{fried 1 F}
		\\
		\big(&2 \Dot{\jfH{}} +3\jfH{}^2 \big)\jfF{} = -\frac{1}{\Planck^2} \left( \frac{1}{2} \Dot{\jfpsi{}}^2 - \jfU{}(\jfpsi{}) \right)\nonumber\\
		&\quad+\alpha \left(-2\jfH{} \Dot{\jfpsi{}} \jfF{}^{\prime} -\Dot{\jfpsi{}}^2 \jfF{}^{\prime \prime} -\jfF{}^{\prime} \Ddot{\jfpsi{}} +\frac{3}{2} \Dot{\jfpsi{}}^2 \jfomega{}^{\prime 2} \right) , 
		\label{fried 2 F}
		\\
		\Ddot{\jfpsi{}}& +3\jfH{} \Dot{\jfpsi{}} +\jfU{}^{\prime}(\jfpsi{}) = \alpha \Big(3 \Planck^2 \jfF{}^{\prime} \left( 2\jfH{}^2 +\Dot{\jfH{}} \right)\nonumber\\
		&\quad+9 \jfH{} \Dot{\jfpsi{}}^2 {\jfomega{}^{\prime}} ^2 +3 \Dot{\jfpsi{}} \jfomega{}^{ \prime 2} \Ddot{\jfpsi{}} +3 \Dot{\jfpsi{}}^3 \jfomega{}^{\prime} \jfomega{}^{\prime \prime} \Big)  . 
		\label{kg F}
	\end{align}
\end{subequations}

From a comparison of  \crefrange{fried 1 F}{kg F} and \crefrange{fried 1 cusc}{kg cusc} it is clear that the field equations are equivalent, with the appropriate choice of $\jfF{}(\jfpsi{}) = 1+\left(\alpha \jfomega{}(\jfpsi{})^2 /2 \right)$. The Lagrangian can be further simplified to a standard scalar-tensor gravity form by substituting for $\jfpsi{}\equiv\jfpsi{}(\jfomega{})$, i.e. the inverse of \cref{omega eq}
\begin{align}\label{xi to omega redef}
	\jfpsi{}(\jfomega{}) \equiv \sqrt{\frac{3}{2}} \Planck \mathrm{arccosh} \left( \frac{\jfomega{}^2 +5 \alpha}{3 \alpha} \right) .
\end{align}
This field redefinition from $\jfpsi{}$ to $\jfomega{}$ means that the $\jfomega{}$ field will inherit the $^\prime$ superscript notation from $\jfpsi{}$, i.e $^\prime$ will now be used to denote a derivate w.r.t the $\jfomega{}$ field.

At this point it is useful to also recall the sign $\alpha$. From~\cref{AlphaDefinition} we notice
\begin{equation}\label{NewAlphaDefinition}
	\alpha=
	\begin{cases}
		-1, & \jfomega{}<0,\\
		1, & 0>\jfomega{}.
	\end{cases}
\end{equation}
This gives the $\jfomega{}$ field an injective correspondence with the $\jfpsi{}$ field (and through the $\jfpsi{}$ field to the corresponding root torsion theory of the scalar-tensor analogue).

This substitution then reduces the Lagrangian to the generalised form of a scalar field non-minimally coupled to the Ricci scalar, in which $\jfomega{}$ carries the single extra dynamical d.o.f
\begin{equation}\label{final lag}
	\jfL{CTEG} =- \frac{\jfF{}(\jfomega{})\Planck ^2}{2} \jfR{} + \frac{\jfB{}(\jfomega{})}{2} \jfX{\jfomega{}\jfomega{}}  
	-\jfU{}(\jfomega{}) +\jfL{M} , 
\end{equation}
with $\jfX{\jfomega{} \jfomega{}} \equiv \frac{\Planck^2}{2} \jfg{^{\mu \nu}} \partial_{\nu} \jfomega{} \partial_{\mu} \jfomega{}$ and the functions
\begin{equation}
	\jfF{}(\jfomega{}) \equiv 1+\alpha \frac{\jfomega{} ^2}{2}, \quad \jfB{}(\jfomega{}) \equiv -\frac{3 \left(16 \alpha +8 \jfomega{}^2 +\alpha \jfomega{}^4 \right)}{ \left(\alpha \jfomega{} ^2+2\right) \left(\alpha \jfomega{} ^2+8\right)} .
\end{equation}
At this point we will assume that a function, unless otherwise stated is a function of $\jfomega{}$, so the $\jfomega{}$ dependence of $\jfF{}=\jfF{}(\jfomega{})$ is implicit.

The field equations for this Lagrangian \cref{final lag} are, upon variation w.r.t the metric and $\jfomega{}$ field
\begin{subequations}	
	\begin{align}
		3&\jfH{}^2 \jfF{} = \frac{\jfB{} \Dot{\jfomega{}}^2}{2} +\jfU{} -3\alpha \jfH{} \jfomega{} \Dot{\jfomega{}} +\jfrho{} ,
		\label{fried 1 final}
		\\
		\big(&3\jfH{}^2+2\Dot{\jfH{}} \big)\jfF{} = -\frac{\jfB{} \Dot{\jfomega{}}^2}{2} +\jfU{} 
		\nonumber\\
		&\quad+\alpha \left(-2\jfH{} \jfomega{} \Dot{\jfomega{}} -\Dot{\jfomega{}}^2 -\jfomega{} \Ddot{\jfomega{}} \right) -\jfP{},
		\label{fried 2 final}
		\\
		\jfB{}& \Ddot{\jfomega{}} +3\jfH{}\jfB{} \Dot{\jfomega{}} +\frac{\jfU{}'}{2} = 3 \alpha \Planck^2 \jfomega{} \left(2\jfH{}^2 +\Dot{\jfH{}} \right)
		\nonumber\\ 
		&\quad-\frac{\jfB{}' \Dot{\jfomega{}}^2}{2} ,
		\label{kg final}
	\end{align}
\end{subequations}
where we have used $\jfrho{}$ and $\jfP{}$ as the energy density and pressure for normal matter. For dust $\jfP{} = 0$, and the continuity equation is
\begin{subequations}
	\begin{equation}
		\Dot{{\hat{\rho}}}_{\text{M}} +3\jfH{}\jfrho{} = 0 .
	\end{equation}
\end{subequations}
The field equations are now in the form where the dynamics of the system can be studied and compared with the literature. For ease of calculation, and as it fits with the choice of most dynamical systems analysis in cosmology, we set the e.o.s for all matter to $w=0$; this choice is purely arbitrary and further analysis can be easily extended to include matter and radiation separately.

\section{Dynamical systems analysis and stability}\label{sec: dyn sys in JF}

\subsection{Constructing the phase space}\label{sec:constructing phase space in JF}

To be able to analyse the dynamics of the system, and the nature of the critical points, we will employ dynamical systems analysis. This is particularly important in showing that the CS is a stable point.

Following the standard method of dynamical systems applied to cosmology \cite{bahamonde2018dynamical}, we write the Friedmann equations \cref{fried 1 final,fried 2 final} in the following form
\begin{subequations}	
	\begin{align}
		1& =  \frac{ \jfB{} \Dot{\jfomega{}}^2}{6\jfH{}^2 \jfF{}} +\frac{\jfU{}}{3\jfH{}^2 \jfF{}}   -\alpha \frac{\jfomega{} \Dot{\jfomega{}}}{\jfH{}\jfF{}}
		+\frac{\jfrho{}}{3\jfH{}^2 \jfF{}}  ,
		\label{dyn rho}	
		\\
		1&+\frac{2 \Dot{\jfH{}}}{\jfH{}^2} =  -\frac{ \jfB{} \Dot{\jfomega{}}^2}{6\jfH{}^2 \jfF{}} +\frac{\jfU{}}{3\jfH{}^2 \jfF{}}   +\alpha \Big( - \frac{2\jfomega{} \Dot{\jfomega{}}}{3\jfH{}\jfF{}} 
		\nonumber\\
		&\quad- \frac{\Dot{\jfomega{}}^2}{3\jfH{}^2 \jfF{}} -\frac{\jfomega{} \Ddot{\jfomega{}}}{3\jfH{}^2 \jfF{}} \Big) ,
		\label{dyn p}
	\end{align}
\end{subequations}
where $\jfrho{} $ represents the energy density of the barotropic matter. \cref{dyn rho} is the Friedmann energy constraint equation, and we will introduce the relative energy densities for dust and the scalar field $\jfomega{}$ as 
\begin{align}\label{e dens}
	\jfOmegaM{} \equiv \frac{\jfrho{} }{3\jfH{}^2 \jfF{}},\quad
	\jfOmegaomega{} \equiv \frac{ \jfB{} \Dot{\jfomega{}}^2}{6\jfH{}^2 \jfF{}} +\frac{\jfU{}}{3\jfH{}^2 \jfF{}}   -\alpha \frac{\jfomega{} \Dot{\jfomega{}}}{\jfH{}\jfF{}}.
\end{align}
With these definitions of the relative energy densities in \cref{e dens}, the Friedmann constraint, reads
\begin{equation}\label{fried Omega constr}
	1 = \jfOmegaM{} +\jfOmegaomega{},
\end{equation}
by assuming $0 \leq \jfOmegaM{} \leq 1$, or equivalently that
\begin{equation}
	0 \leq \jfOmegaomega{} \leq 1.
\end{equation}
Denoting $'$ as the derivative $\mathrm{d/dN}$ being the derivative w.r.t e-folds $N \equiv \ln(a)$ (a note for clarity, this $^\prime$ notation only applies for the phase space variables $\jfx{},\jfy{},\jfz{}$), and choosing the dynamical variables
\begin{align}\label{JF dyn vars}
	\jfx{} \equiv \frac{\Dot{\jfomega{}}}{\jfH{}}, \quad \jfy{} \equiv \frac{1}{\jfH{}} \sqrt{\frac{\jfU{}}{3\jfF{}}}, \quad \jfz{} \equiv \jfomega{},
\end{align}
the Friedmann constraint \cref{fried Omega constr} reduces to
\begin{equation}\label{phase space constraint}
	0\leq \jfx{}^2 \frac{\jfB{}}{6\jfF{}} +\jfy{}^2 -\alpha \frac{\jfx{}\jfz{}}{\jfF{}} \leq 1 .
\end{equation}
Note that in this work the construction of the phase space for the theory is the main aim, and we neglect considerations of energy conditions in scalar field cosmology, in particular with regards to phantom scalar fields. We restrict ourselves to the construction of the phase space, as this is the most concise way to analyse the dynamical properties of the system. The phase space region is given by $\jfomega{} > -\sqrt{2}$; this lower bound is from the breakdown of the theory as the prefactor in front of the Ricci scalar, $\jfF{} =1+\left(\alpha \jfomega{}^2/2\right)$, vanishes as $\jfomega{} \rightarrow -\sqrt{2}$.

To construct the dynamical system from the dynamical variables, it is necessary to manipulate the system into the standard dynamical systems form
\begin{equation}\label{x=mx}
	\textbf{X}^\prime \left(N\right)   = \mathbf{F}\left(\textbf{X}\left(N\right)\right) = \mathbf{F}\left(\textbf{X}\right),
\end{equation}
where $\textbf{X}$ is a state vector of the system dependent on a single variable $N$ (number of e-folds), and $\textbf{F}$ is a vector field. This is the defining characteristic of an autonomous ODE: the system of ODEs doesn't have an explicit dependence on $N$. This means that the stationary solutions are time invariant, if a system started at a point in the phase space $\textbf{X}_0$, such that $\textbf{X}^\prime _0 (N)   = \mathbf{F}\left(\textbf{X} _0\right) = 0$, then the system would stay at the point $\textbf{X}_0$ for any transformation in $N$, such as $N \rightarrow N + \delta N$. 

By substituting the dynamical variables \cref{JF dyn vars} into \crefrange{dyn rho}{dyn p} we obtain  
\begin{subequations}
	\begin{widetext}
		\begin{align}
			\jfx{}'(N) &= \frac{3 \left(\jfB{}^2 \jfx{}^3+\jfB{} \jfx{} \left(\alpha  \jfx{} (2 \jfx{}-5 \jfz{})-6 \jfF{} \left(\jfy{}^2+1\right)\right)+6 \alpha  \jfz{} \left(3 \jfF{} \jfy{}^2+\jfF{}-\alpha  \jfx{}^2\right)\right)-(2 \jfF{}+\alpha  \jfx{} \jfz{}) \left(3 \jfB{} \jfx{}^2 \lambda_{\jfB{}}+\jfy{}^2 \lambda_{\jfU{}} \right) }{6 \left(2 \jfB{} \jfF{}+3 \alpha ^2 \jfz{}^2\right)}{6 \left(2 \jfB{} \jfF{}+3 \alpha ^2 \jfz{}^2\right)}, \label{x' eq JF} \\
			\jfy{}'(N) &= \frac{\jfy{}}{6} \left[\frac{3 \jfB{} \left(\jfB{} \jfx{}^2-6 \jfF{} \left(\jfy{}^2-1\right)\right)-\alpha  \jfz{} \left(3 \jfB{} \jfx{}^2 \lambda_{\jfB{}}+\jfy{}^2 \lambda_{\jfU{}} \right)+6 \alpha  \jfB{} \jfx{} (\jfx{}-\jfz{})+36 \alpha ^2 \jfz{}^2}{2 \jfB{} \jfF{}+3 \alpha ^2 \jfz{}^2}-\frac{3 \alpha  \jfx{} \jfz{}}{\jfF{}}+3 \jfx{} \lambda_{\jfU{}}\right], \label{y'eq JF} \\
			\jfz{}'(N) &= \jfx{} ,\label{z' eq JF}
		\end{align}
	\end{widetext}
\end{subequations}

with $\lambda_{\jfU{}} \equiv \frac{1}{\jfU{}} \frac{\mathrm{d}\jfU{}}{\mathrm{d}\jfomega{}}$ and $\lambda_\jfB{} \equiv \frac{1}{\jfB{}} \frac{\mathrm{d}\jfB{}}{\mathrm{d}\jfomega{}}$. The equations \crefrange{x' eq JF}{z' eq JF} form a closed dynamical system.

\subsection{Stability and study of critical points}\label{sec: stability in JF}
\begin{figure*}[t]
	\centering
	\includegraphics[width=\textwidth]{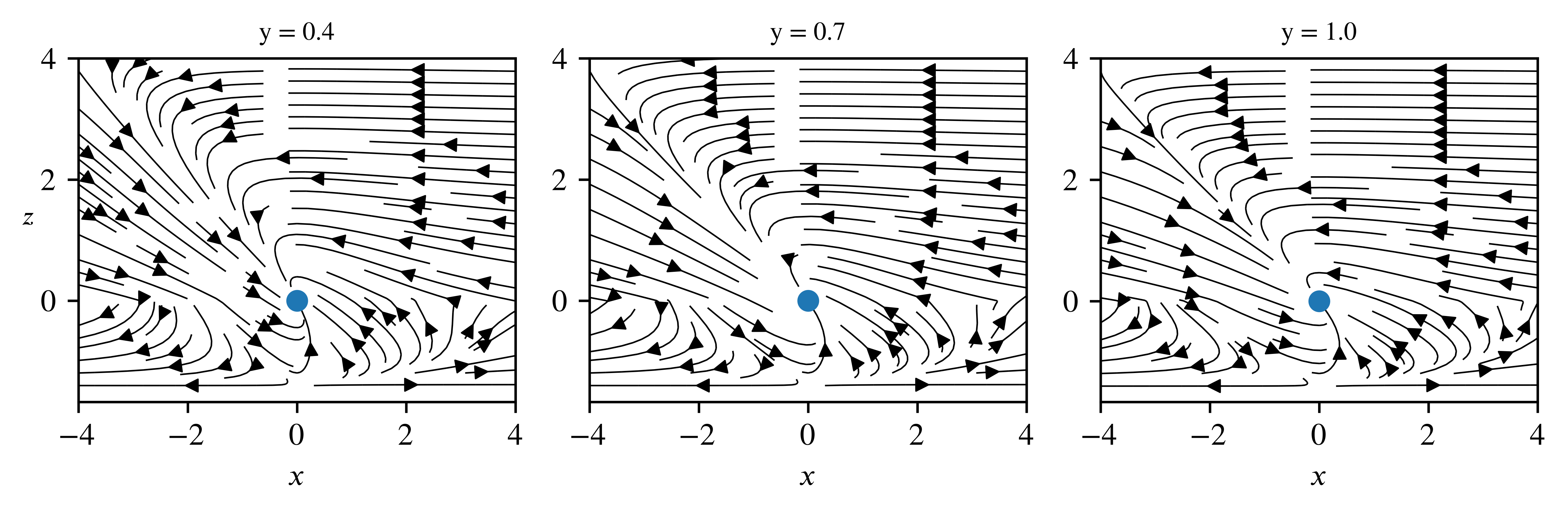}
	\caption{Plot of phase space of system \crefrange{x' eq JF}{z' eq JF}, the blue marker is the CS point at $\left(\jfx{},\jfy{},\jfz{}\right) = \left(0,1,0\right)$. The plots are slices of the $\jfx{}-\jfz{}$ plane, with constant $\jfy{}$}
	\label{fig:phase space JF with critical points}
\end{figure*}

To study the critical points of this system, we employ the use of linear stability theory (see \cite{wiggins2003introduction}, or one of the many review articles on dynamical systems applied to cosmology \cite{bahamonde2018dynamical}). We start by defining 
\begin{equation}\label{f(x_c)}
	f(\mathbf{\jfx{}}) \equiv \begin{pmatrix}
		P \\
		Q \\
		R
	\end{pmatrix} , 
\end{equation}
and the Jacobian (or stability matrix) \cite{bahamonde2018dynamical} of the system to be 
\begin{equation}\label{jacobian}
	J \equiv \begin{pmatrix}
		\partial_\jfx{} P && \partial_\jfy{} P && \partial_\jfz{} P \\
		\partial_\jfx{} Q && \partial_\jfy{} Q && \partial_\jfz{} Q\\
		\partial_\jfx{} R && \partial_\jfy{} R && \partial_\jfz{} R 
	\end{pmatrix} ,
\end{equation}
where we have defined $P \equiv \jfx{}'$, $Q \equiv \jfy{}'$ and $R \equiv \jfz{}'$ from \crefrange{x' eq JF}{z' eq JF}.

To identify the critical points of the system we must find the points $\mathbf{\jfx{}_c}$ that satisfy $f\left(\mathbf{\jfx{}_c}\right) = 0 $ from \cref{f(x_c)}. For the range of $\jfomega{} > -\sqrt{2}$ the critical points are 
\begin{align}\label{crit points}
	\mathbf{\jfx{}_{\pm}} \equiv \begin{pmatrix}
		0 \\
		\pm 1 \\
		0
	\end{pmatrix}, \quad \mathbf{\jfx{}_2} \equiv \begin{pmatrix}
		0 \\
		0 \\
		0
	\end{pmatrix}  . 
\end{align}
The CS point is to be found at $f\left(\mathbf{\jfx{}_{\pm}}\right)$ (note the $\pm$ is from the fact that variable $\jfy{}$ is defined as the square root of the potential: the dynamics are the same for both signs we assume that the potential is always real). To analyse the stability of  the correspondence solution, the eigenvalues of the Jacobian \cref{jacobian} need to be evaluated at the point $\mathbf{\jfx{}_{\pm}}$. In \cref{eigenval of cs} it can be seen that all the eigenvalues of $J\left(\mathbf{\jfx{}}_{\pm}\right)$ have negative real parts \cite{bahamonde2018dynamical}
\begin{equation}\label{eigenval of cs}
	J(\mathbf{\jfx{}}_\pm) = \begin{pmatrix}
		-3 \\
		\frac{1}{12} (-18 - \sqrt{46}) \\
		\frac{1}{12} (-18 + \sqrt{46}) 
	\end{pmatrix} .
\end{equation}
Thus, we find that the CS point in the phase space, corresponding to $\jfx{}=0$, $\jfy{}=1$ and $\jfz{}=0$, is a stable point. Translating this back into the physical quantities of the $\jfomega{}$ scalar field, with the definitions in \cref{JF dyn vars}, the first Friedmann equation reads
\begin{align}
	3\jfH{}^2 = \jfx{}^2 \frac{\jfB{}}{6\jfF{}} +\jfy{}^2-\frac{\alpha \jfz{} \jfx{}}{\jfF{}} +\jfrho{} =  \lambda +\jfrho{}.
\end{align}
This shows that the late time behaviour of the theory is exactly that of GR with a cosmological constant. With $\jfy{} = \pm 1$ and the other terms being zero, the potential has found its minimum value $\jfomega{} =0$ at the CS, and thus the only quantity left in the Friedmann equations from the $\jfomega{}$ field is the $\lambda$ constant which mimics the role of the cosmological constant of \LCDM{} cosmology. This is the motivation for referring to the $\lambda$ coupling in CTEG~\cref{CTEG} as an emergent cosmological constant, we recall that the true \emph{external} cosmological constant $\Lambda$ is not yet included in the model, but will be re-introduced in~\cref{ReintroduceLambda}.

This confirms the findings of \cite{Barker_2020}, in which the correspondence solution was only studied graphically, but in our formulation of the Lagrangian it is possible to use the rigorous toolkit of linear stability analysis to study the correspondence solution.  

The dynamics of the phase space can be presented in a pictorial manner by using a phase space plot, as in \cref{fig:phase space JF with critical points}. This plot shows the flow of the autonomous system \crefrange{x' eq JF}{z' eq JF}. As can be seen in \cref{fig:phase space JF with critical points} the point $\mathbf{\jfx{}_{\pm}}$ acts as an asymptotically stable sink point, this being the point marked in blue, and denotes the correspondence solution at which the theory equates to GR with a cosmological constant. 

The CS solution, $\mathbf{\jfx{}}_{\pm}$ is classified as a stable node point, as the eigenvalues of the Jacobian \cref{jacobian} are all negative real parts. The critical point $f\left(\mathbf{\jfx{}}_2\right)$ is an unstable node.

\section{Einstein frame dynamics}\label{sec:EF}
We delineate between quantities in the Jordan frame of the preceding section, denoted with a hat, and quantities defined in the Einstein frame using a tilde. 

To move from the Jordan frame to the Einstein frame, we must perform a conformal transformation \cite{faraoni2004cosmology}. This conformal transformation will take the form of $g_{\mu \nu}=\efOmega{}(\jfomega{})^2 \efg{_{\mu\nu}}$ relative to the original (physical) frame of~\cref{CTEG}, such that through its definition relative to the preceding frame it can remove the curvature prefactor
\begin{align}\label{conformal transformation}
	\frac{\efOmega{}(\jfomega{})^2}{\jfOmega{}(\jfomega{})^2} = \jfF{}(\jfomega{}) = 1+ \left(\alpha \jfomega{} ^2 /2 \right).
\end{align}
This transformation will take the Lagrangian \cref{final lag} from the Jordan frame to the Einstein frame Lagrangian
\begin{equation}\label{Einstein frame lagrangian}
	\efL{CTEG} = \Planck ^2 \left( - \frac{1}{2} \efR{} +2\efB{}(\jfomega{}) \efX{\jfomega{}\jfomega{}}\right) -\efU{}(\jfomega{}) +\efL{M},
\end{equation}
with $\efX{\jfomega{}\jfomega{}}\equiv\frac{1}{2}\efg{^{\mu\nu}}\partial_\mu\jfomega{}\partial_\nu\jfomega{}$, and $\efU{}(\jfomega{})$ denoting the potential in the Einstein frame, and $\efB{}(\jfomega{})$ representing the new non-canonical factor in front of the kinetic term. It should be noted that the matter Lagrangian $\efL{M}$ will have picked up a further coupling to $\jfomega{}$ through the conformal transformation \cref{conformal transformation}; this coupling will be from the two conformal transformations that have been made, with the first from~\cref{JFOmega}, and the second from \cref{conformal transformation}, thus giving the combined coupling of
\begin{equation}
	\efOmega{}(\jfomega{})^2=\left(1+\left(\alpha \jfomega{}^2/2 \right) \right) \left(1+\left(\alpha \jfomega{}^2/8 \right) \right).  
\end{equation}
Meanwhile, $\efB{}(\jfomega{})$ and $\efU{}(\jfomega{})$ are related to the original $\jfB{}(\jfomega{})$ and $\jfU{}(\jfomega{})$ by 
\begin{subequations}
	\begin{align}
		\efU{}(\jfomega{}) &= \frac{\jfU{}(\jfomega{})}{\jfF{}(\jfomega{})^2} =\lambda \left(  \frac{3}{2 \alpha  \jfomega{} ^2+4}+\frac{1}{4} \right),\\ 
		\efB{}(\jfomega{})&= 2\left(\frac{\jfB{}(\jfomega{})}{2\jfF{}(\jfomega{})} +\frac{3 \jfomega{}^2}{4\jfF{}(\jfomega{})^2}\right) 
		\nonumber
		\\
		&= -\frac{96 \alpha}{(2+\alpha \jfomega{}^2)^2 (8+\alpha \jfomega{}^2)}.
	\end{align}
\end{subequations}
The field equations for the Lagrangian \cref{Einstein frame lagrangian} are 
\begin{subequations}
	\begin{align}		
		3\efH{} ^2 &= \frac{1}{2}\efB{} \Dot{\jfomega{}} ^2 +\efU{} +\efrho{},
		\label{fried 1 EF}
		\\
		3\efH{} ^2 + 2 \Dot{\efH{}} &= -\frac{1}{2}\efB{} \Dot{\jfomega{}} ^2 +\efU{} ,
		\label{fried 2 EF}
		\\
		\efB{} \Ddot{\jfomega{}} + 3\efH{} \efB{} \Dot{\jfomega{}} +\efU{}^\prime &= -\frac{1}{2} \efB{} ^\prime \Dot{\jfomega{}}^2,
		\label{KG EF}
	\end{align}
\end{subequations}
where the superscript $^\prime$ denotes a derivative w.r.t the $\jfomega{}$ field, i.e $\mathrm{d}/\mathrm{d} \jfomega{}$.

\subsection{Dynamical systems analysis in the Einstein frame}\label{sec: dyn sys EF}

From the field equations \crefrange{fried 1 EF}{KG EF} a dynamical systems analysis similar to the analysis in the Jordan frame can be performed. First, we  divide \crefrange{fried 1 EF}{fried 2 EF} by $3\efH{} ^2$ to get the following set of equations
\begin{subequations}
	\begin{align}	
		1 &= \frac{\efB{} \Dot{\jfomega{}} ^2}{6 \efH{} ^2} +\frac{\efU{}}{3\efH{} ^2} +\frac{\efrho{}}{3\efH{} ^2} , 
		\label{1=... fried 1 EF}
		\\	
		1+\frac{2}{3} \frac{\Dot{\efH{}}}{\efH{} ^2} &= - \frac{\efB{} \Dot{\jfomega{}} ^2}{6 \efH{} ^2} +\frac{\efU{}}{3\efH{} ^2} .
		\label{1=... fried 2 EF}
	\end{align}	
\end{subequations}
From this set of equations, we can define the set of dynamical variables that will describe the evolution of the system
\begin{equation}\label{EF dyn vars}
	\efx{} \equiv \frac{\Dot{\jfomega{}}}{\efH{}}, \quad \efy{} \equiv \frac{1}{\efH{}} \sqrt{\frac{\efU{}}{3}}, \quad \efz{} \equiv \jfomega{}.
\end{equation}
The Friedmann constraint \cref{fried 1 EF}, with the definition of the relative matter density $ \efOmegaM{} \equiv \efrho{}  / 3\efH{} ^2$, becomes 
\begin{equation}
	1 - \efOmegaM{} = \frac{\efB{}}{6} \efx{} ^2 +\efy{} ^2 .
\end{equation}
Using these dynamical variables, the phase space can be set up, as in the previous section detailing the use of dynamical systems techniques in the Jordan frame, by getting the equations into the form in \cref{x=mx}. The set of autonomous differential equations describing \crefrange{1=... fried 1 EF}{1=... fried 2 EF} is given by 

\begin{subequations}
	\begin{widetext}
		\begin{align}
			\efx{} '(N) & = -\frac{24 \alpha  \efx{} ^3}{\left(\alpha  \efz{} ^2+2\right)^2 \left(\alpha  \efz{} ^2+8\right)}+\frac{3 \alpha  \efx{} ^2 \efz{}  \left(\alpha  \efz{} ^2+6\right)}{\left(\alpha  \efz{} ^2+2\right)
				\left(\alpha  \efz{}^2+8\right)}-\frac{3}{2} \efx{}  \left(\efy{} ^2+1\right)-\frac{3}{8} \efy{} ^2 \efz{}  \left(\alpha  \efz{} ^2+2\right), \label{x' eq EF} \\
			\efy{} '(N) & = \frac{3}{2} \efy{}  \left(-\frac{4 \alpha  \efx{}  \left(4 \efx{} +\alpha  \efz{} ^3+2 \efz{} \right)}{\left(\alpha  \efz{} ^2+2\right)^2 \left(\alpha  \efz{} ^2+8\right)}-\efy{} ^2+1\right), \label{y'eq EF} \\
			\efz{} '(N) & = \efx{}  . \label{z' eq EF} 
		\end{align}
	\end{widetext}
\end{subequations}

\subsection{Stability analysis and critical points in the Einstein frame}\label{sec:stability in EF}

Now that the field equations resulting from the variation of the Lagrangian \cref{Einstein frame lagrangian} w.r.t the metric and the scalar field $\jfomega{}$ are in the form of an autonomous dynamical system \crefrange{x' eq EF}{z' eq EF} we can move on to the study of the critical points and stability of the system. 

Following the stability analysis in \cref{sec: dyn sys in JF}, we start with the definition of the Jacobian \cref{jacobian} to find the critical points, which read
\begin{align}\label{crit points EF}
	\mathbf{\efx{} ^{\pm}} \equiv \begin{pmatrix}
		0 \\
		\pm 1 \\
		0
	\end{pmatrix},  \quad \mathbf{\efx{} ^2} \equiv \begin{pmatrix}
		0 \\
		0 \\
		0
	\end{pmatrix} ,
\end{align}
The nature of the stability of the $\mathbf{\efx{} ^{\pm}}$ point can be determined by calculating the eingenvalues of the Jacobian for the system evaluated at $\mathbf{x} = \mathbf{\efx{} ^{\pm}}$. This results in 
\begin{equation}\label{eigenval of cs EF}
	J(\mathbf{\efx{} ^{\pm}}) =
	\begin{pmatrix}
		-3 \\
		\frac{1}{2} (-3 - \sqrt{6}) \\
		\frac{1}{2} (-3 + \sqrt{6})  
	\end{pmatrix},
\end{equation}
where again the CS point features all negative real parts of the eigenvalues of the Jacobian, and no imaginary parts, thus satisfying the stability condition of linear stability theory and showing mathematically that the CS is a feature of both the Einstein and Jordan frame Lagrangia. 

For the point $\mathbf{\efx{} ^{\pm}}$ the eigenvalues of the Jacobian are 
\begin{equation}\label{eigenval of repeller EF}
	J(\mathbf{\efx{} ^{\pm}}) = \frac{3}{2} \begin{pmatrix}
		-1 \\
		1 \\
		0
	\end{pmatrix} ,
\end{equation}
making this point an unstable point in the phase space.

An interesting point to note is that the stability in both the Einstein and Jordan frames is independent of $\alpha$; this can be seen by the lack of the $\alpha$ terms in any of the eigenvalues of the above stability analyses. This is something that the original work on the scalar-tensor analogue \cite{Barker_2020} was unable to show, as that analysis was (erroneously) restricted to values of torsion above the correspondence solution. 

\begin{figure*}[t]
	\centering
	\includegraphics{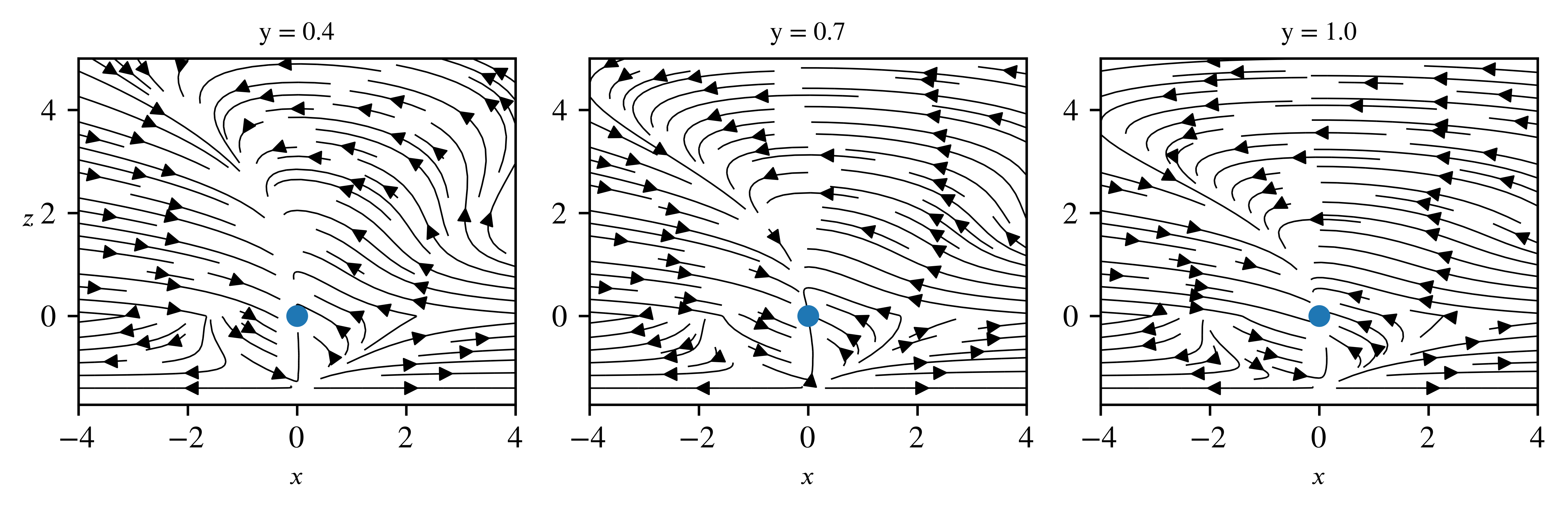}
	\caption{Plots of the Einstein frame phase space, wit the blue marker representing the CS (correspondence solution), with the slices of the phase space being in the $x-z$ plane at constant $y$ values indicated above each slice.}
	\label{fig:EF phase space}
\end{figure*}

\section{Inflationary applications}\label{sec:inflation}

\subsection{Canonical single scalar field in the Einstein frame}\label{sec:field redefinition}

In this section we take a brief look at the inflationary cosmology associated with the scalar-tensor analogue, by performing a field redefinition of the form $\jfomega{}\equiv\jfomega{}(\efpsi{})$, where
\begin{align}\label{field redefinition}
	\efpsi{}(\jfomega{}) \equiv 4\int\mathrm{d} \jfomega{} \sqrt{\left| -\frac{6\alpha}{(2+\alpha \jfomega{}^2)^2 (8+\alpha \jfomega{}^2)} \right|},
\end{align}
and the moduli have been included as the scalar field is a phantom for $\jfomega{}>0$.
The solution to~\cref{field redefinition} can be found analytically, and inverted, so that the potential $\efU{}(\efpsi{})$ can be written analytically as a function of just $\efpsi{}$. This reduces the system to a surprisingly simple form
\begin{widetext}
	\begin{equation}\label{cosh and cos lag}
		\efL{CTEG}=
		\begin{cases}
			-\frac{\Planck^2}{2}\efR{}+\efX{\efpsi{}\efpsi{}}-\efU{}(\efpsi{})+\efL{M}
			\vphantom{\left(\frac{\efpsi{}}{\sqrt{8}}\right)},\\
			\quad\\
			-\frac{\Planck^2}{2}\efR{}-\efX{\efpsi{}\efpsi{}}-\efU{}(\efpsi{})+\efL{M}
			\vphantom{\left(\frac{\efpsi{}}{\sqrt{8}}\right)},
		\end{cases}
		\quad
		\efU{}(\efpsi{})=
		\begin{cases}
			\Planck^2
			\lambda\cosh^2\left(\frac{\efpsi{}}{\sqrt{8}}\right), &\efpsi{}\leq 0,\\
			\quad\\
			\Planck^2
			\lambda\cos^2\left(\frac{\efpsi{}}{\sqrt{8}}\right), & 0<\efpsi{}<\frac{\sqrt{8}\pi}{3},
		\end{cases}
	\end{equation}
\end{widetext}
with $\efX{\efpsi{}\efpsi{}}\equiv\frac{\Planck^2}{2}\efg{^{\mu\nu}}\tensor{\partial}{_\mu}\efpsi{}\tensor{\partial}{_\nu}\efpsi{}$.

\begin{figure}[h]
	\centering
	\includegraphics{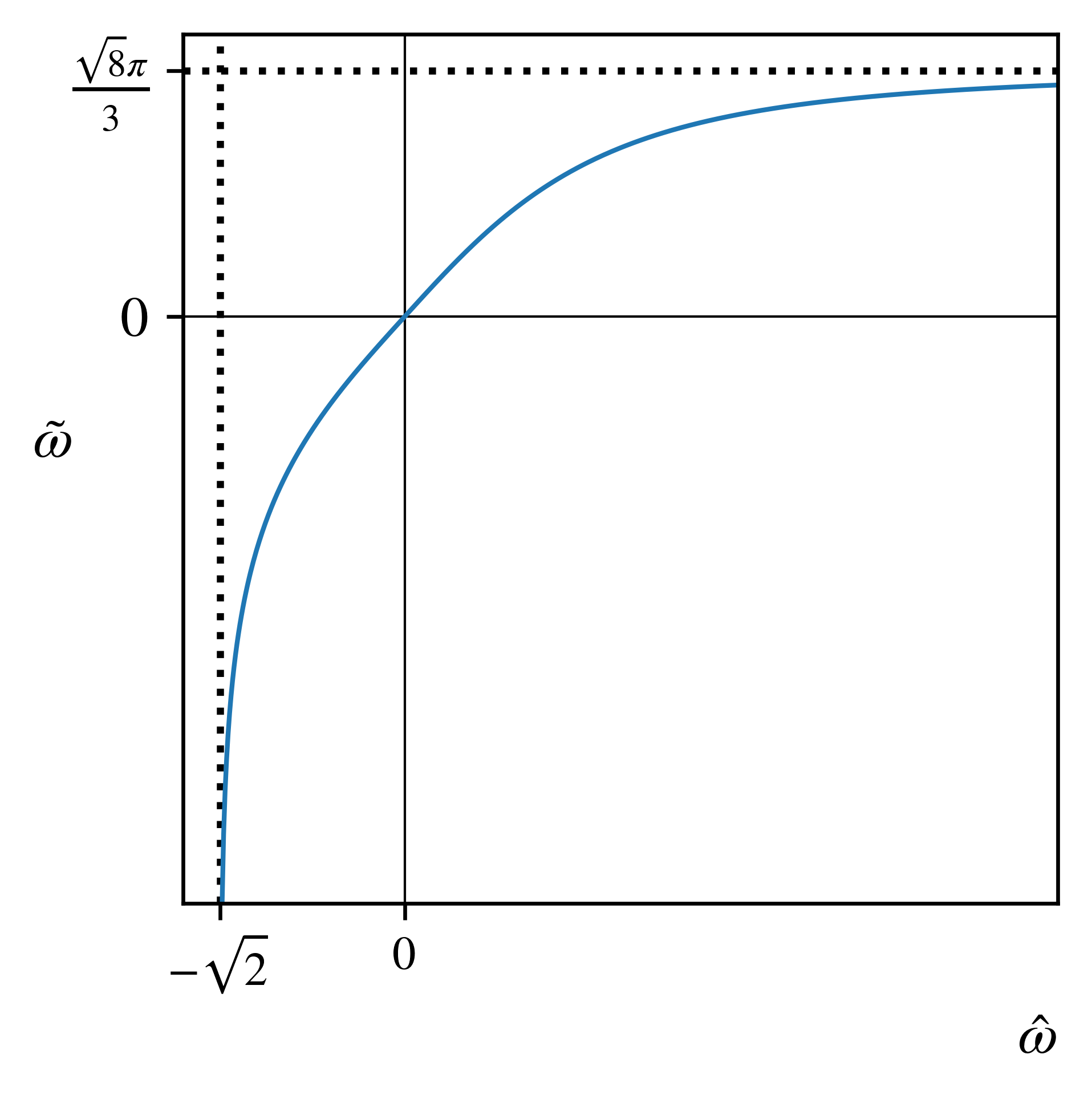}
	\caption{Plot of the solution of the field redefinition \cref{field redefinition}, with the two asymptotic regions show. As the field redefined field passes through the origin the CS point will remain at $\jfomega{}=\efpsi{} =0$.}
	\label{fig:field redefinition plot}
\end{figure}
A plot of the solution found for $\efpsi{}$ is included as it is useful to understand the nature of the system. As can be seen from \cref{fig:field redefinition plot} the $\efpsi{}$ function has 2 distinct regions. One point to note is that the original $\jfomega{}$ field and the redefined $\efpsi{}$ field both have a the same crossing of the origin at $\efpsi{}(\jfomega{} = 0)  = 0$, so the CS point is still at the origin for the redefined field $\efpsi{}$.

In \cref{fig:field redefinition plot} there can be seen two different asymptotic regions of the redefined field $\efpsi{}$. As ${\jfomega{} \rightarrow -\sqrt{2}}$ the prefactor of the Ricci scalar in \cref{final lag} vanishes ${\jfF{}(\jfomega{}) \rightarrow 0}$; this point is seen as the limit ${\efpsi{} \rightarrow - \infty}$. Also, as ${\jfomega{} \rightarrow \infty}$ the re-canonicalised field approaches a finite limit ${\efpsi{} \rightarrow \left(\sqrt{8} \pi \right)/3}$.
\begin{figure}
	\centering
	\includegraphics{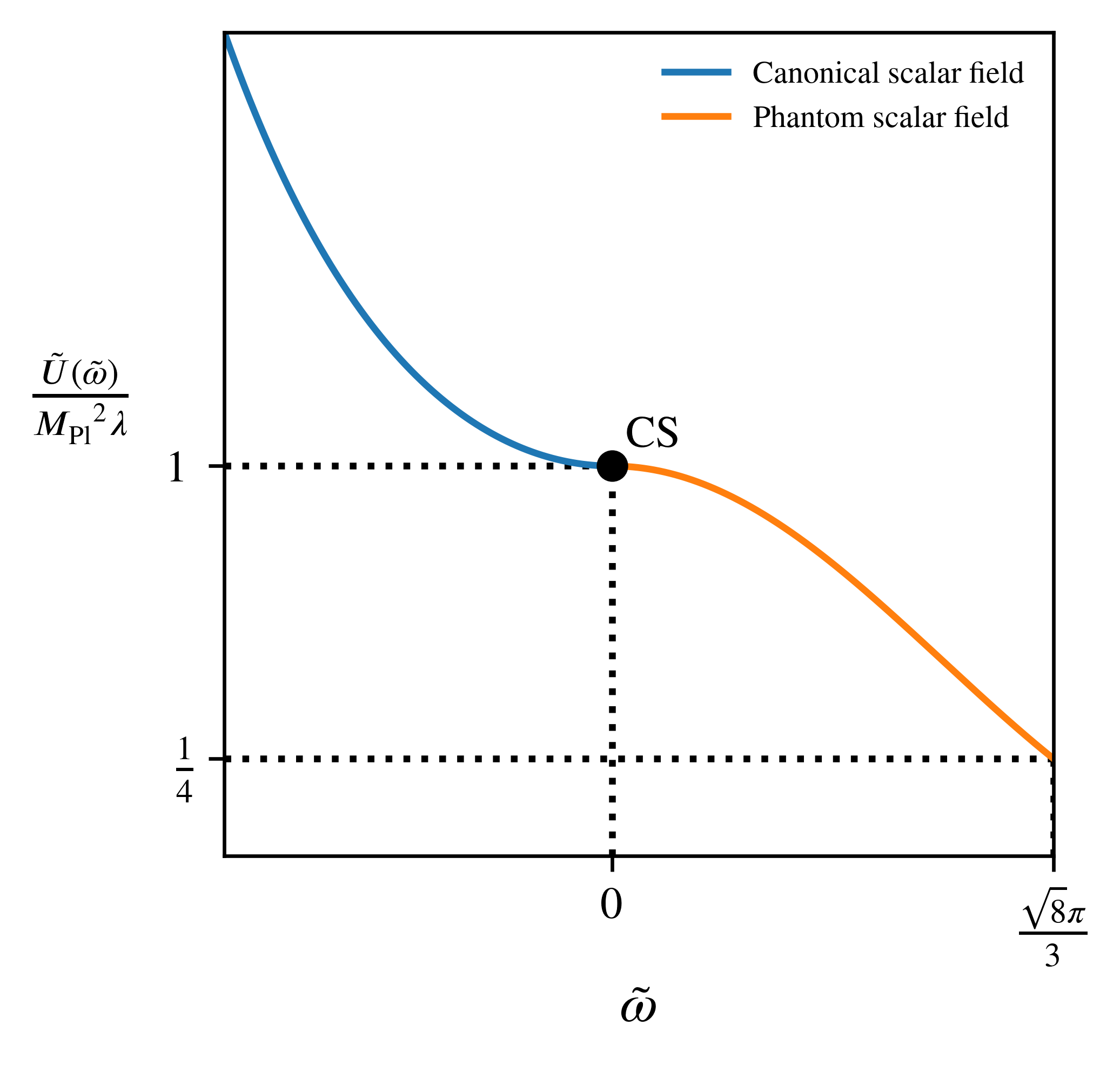}
	\caption{Plot of the potential of the scalar field, with the orange line representing the phantom (`roll up-hill') region, and the blue line representing the standard canonical scalar field region.}
	\label{fig:inflation plot}
\end{figure}

One important feature of the potential in \cref{fig:inflation plot} is that for $\efpsi{} > 0$ the field is a phantom scalar. We expect this to manifest as a field `rolling up' its potential, whereas for $\efpsi{} \leq 0$ the field is a standard canonical scalar field rolling down its potential. 

\subsection{Bare cosmological constant and potential for inflation}\label{ReintroduceLambda}

Now that the system introduced in \cref{cuscuton lag} has been simplified to the form \cref{cosh and cos lag}, we can include a brief discussion of the addition of an external cosmological constant $\Lambda$, and the inflationary implications therein. Note that $\Lambda$ is not a conformal term in the physical frame~\cref{CTEG}, and so it will pick up scalar couplings with each conformal transformation in~\cref{JFOmega,conformal transformation}. Tracing these through, this has the effect of changing the form of the potential $\efU{}$ to 
\begin{align}\label{external cosmo pot}
	\efU{}(\jfomega{}) =  \frac{8 \alpha  \jfomega{} ^2 (5 \lambda +2 \Lambda )+\jfomega{} ^4 (4 \lambda +\Lambda )+64 (\lambda +\Lambda )}{16 \left(\alpha  \jfomega{} ^2+2\right)^2},
\end{align}
The form of the potential $\efU{}(\efpsi{})$, once the field redefinition in \cref{sec:field redefinition} has been applied, reads 
\begin{align}\label{tilde potential}
	\efU{}(\efpsi{}) &= \frac{1}{2}\Planck^2 \cosh ^2\left(\frac{\efpsi{}}{\sqrt{8}}\right) \nonumber\\
	&\quad \times\left[2 \lambda +\Lambda +\Lambda  \cosh \left(\frac{\efpsi{}}{\sqrt{2}}\right)\right] ,
\end{align}
where we have restricted the scope of this section to looking at values of $\efpsi{}\leq 0$.

With this new potential, we can use the dynamical systems framework to find the conditions for stability with the addition of the parameter $\Lambda$. The dynamical systems analysis follows the same method as \cref{sec: dyn sys in JF} and \cref{sec:EF}, and we will not repeat all the steps here. 

The stability condition we impose is the requirement that all the real parts of the eigenvalues for the Jacobian of the dynamical system \cref{jacobian} are less than zero. This condition is ambivalent to the nature of the stable point (i.e spiral or sink), but rather the asymptotic stability of the point. With this in mind, we find that the region of validity for the two parameters $\lambda$ and $\Lambda$ reads
\begin{align}\label{bounds of stability external cosmo const}
	-2 \nleq \frac{\lambda}{\Lambda} \nleq -1 . 
\end{align}

\begin{figure}[t!]
	\centering
	\includegraphics[width=\columnwidth]{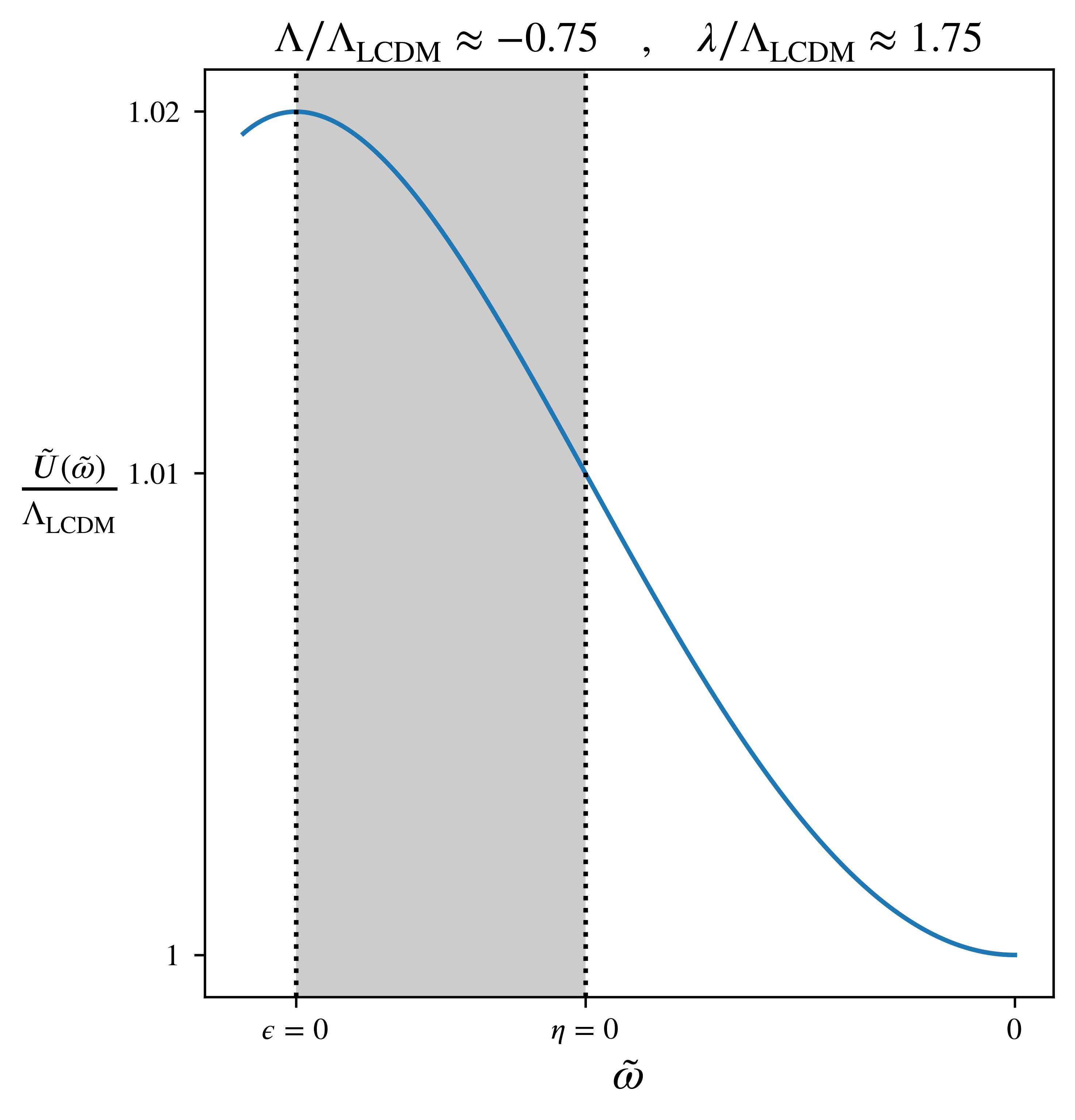}
	\caption{Plot of the potential with the parameters $\Lambda$,$\lambda$ chosen such that there are 50 e-folds of inflation. The shaded region shows the region integrated over in \cref{efold eq}. We have normalised the potential such that the final value of 1 at the CS corresponds to the potential having the value of $\LambdaLCDM{}$.}
	\label{fig:potential for inflation}
\end{figure}

An interesting regime that could serve as an inflationary potential in this theory is for the values of $\Lambda<0$ and $\lambda>0$. This regime has a \emph{negative external cosmological constant}, but allows for the formation of a hilltop in the potential. The slow-roll parameters for inflation are defined by \cite{Liddle_1994}
\begin{subequations}
	\begin{align}
		\epsilon &\equiv \frac{1}{2} \left(\frac{1}{\efU{}(\efpsi{})} \frac{\mathrm{d} \efU{}(\efpsi{})}{\mathrm{d} \efpsi{}} \right) ^2 \quad  \notag\\
		&= \frac{2 \left(\lambda  \tanh \left(\frac{\efpsi{} }{\sqrt{8}}\right)+\Lambda  \sinh \left(\frac{\efpsi{} }{\sqrt{2}}\right)\right)^2}{\left(2 \lambda +\Lambda  \cosh
			\left(\frac{\efpsi{} }{\sqrt{2}}\right)+\Lambda \right)^2},
		\\
		\eta &\equiv \left( \frac{1}{\efU{} (\efpsi{})} \frac{\mathrm{d}^2 \efU{}(\efpsi{})}{\mathrm{d} \efpsi{}^2}  \right) \quad  \notag\\
		&= \Bigg|4 -\left(6 \lambda +5 \Lambda \right)\left[2 \lambda +\Lambda +\Lambda  \cosh \left(\frac{\efpsi{} }{\sqrt{2}}\right)\right]^{-1}\nonumber\\
		&\hspace{80pt}-\left[\cosh \left(\frac{\efpsi{}}{\sqrt{2}}\right)+1\right]^{-1}\Bigg|.
	\end{align}
\end{subequations}
We consider inflation between a point near $\epsilon=0$ at the top of the hill in~\cref{fig:potential for inflation}, and the inflection point at $\eta =0$. This region is stable, as shown in the preceding sections, and evolves towards the CS (note that this stability applies at the background level: the potential is only valid at the background level and the possibility of quantum effects, e.g tunnelling, are not considered). The heuristic constraints placed on the potential include the positive total cosmological constant seen at the CS at $\efpsi{} = 0$, and the stability bounds of the CS given in \cref{bounds of stability external cosmo const}. These two conditions lead to 
\begin{align}\label{sum of lambdas = LCDM}
	\begin{rcases}
		\Lambda + \lambda = \LambdaLCDM{}  \\
		\quad \\
		-2 \nleq \frac{\lambda}{\Lambda} \nleq -1
	\end{rcases} \implies \begin{cases}
		\Lambda= \LambdaLCDM{} -\lambda, \\
		\quad \\
		\LambdaLCDM{} < \lambda <2 \LambdaLCDM{}.
	\end{cases} 
\end{align}
The number of e-folds, $\mathrm{N_e}$ for slow-roll inflation is given in terms of the slow-roll parameter $\epsilon$, and is defined as \cite{Liddle_1994}

\begin{align}\label{efold eq}
	\mathrm{N_{e}} \equiv \int_{t_i}^{t_f} \efH{} \mathrm{d}t \approx \int_{\efpsi{}_i}^{\efpsi{}_f} \frac{\mathrm{d} \efpsi{}}{\sqrt{2 \epsilon}},
\end{align}
where the subscripts $i,f$ denote the beginning and end of inflation respectively. These limits are shown in \cref{fig:potential for inflation} by the shaded region, and they correspond to the top of the hill at $\efpsi{}_i$, and the point where $\mathrm{d} \epsilon / \mathrm{d} \lambda =0$. Note we have added a small shift from the hill top of the form $\efpsi{}+ \delta$, where $\delta$ is a small positive perturbation from the hilltop, so that the field will roll down to the CS point.

From \cite{planck2018}, the value of the cosmological constant is $\LambdaLCDM{} = 2.846 \times 10^{-122} \Planck^2$. To find this in terms of the two parameters of the theory $\Lambda$ and $\lambda$, an application of~\cref{sum of lambdas = LCDM} yields
\begin{subequations}
	\begin{align}
		\lambda &= 1.752 \times \LambdaLCDM{} = 4.987 \times10^{-122} \Planck^2, \\
		\Lambda &= \LambdaLCDM{} -\lambda = -2.141 \times 10^{-122} \Planck^2.
	\end{align}
\end{subequations}
We will return to these values in~\cref{Conclusions}.
\subsection{Inflationary phase space}

\begin{figure*}[t!]
	\centering
	\includegraphics[width=\textwidth]{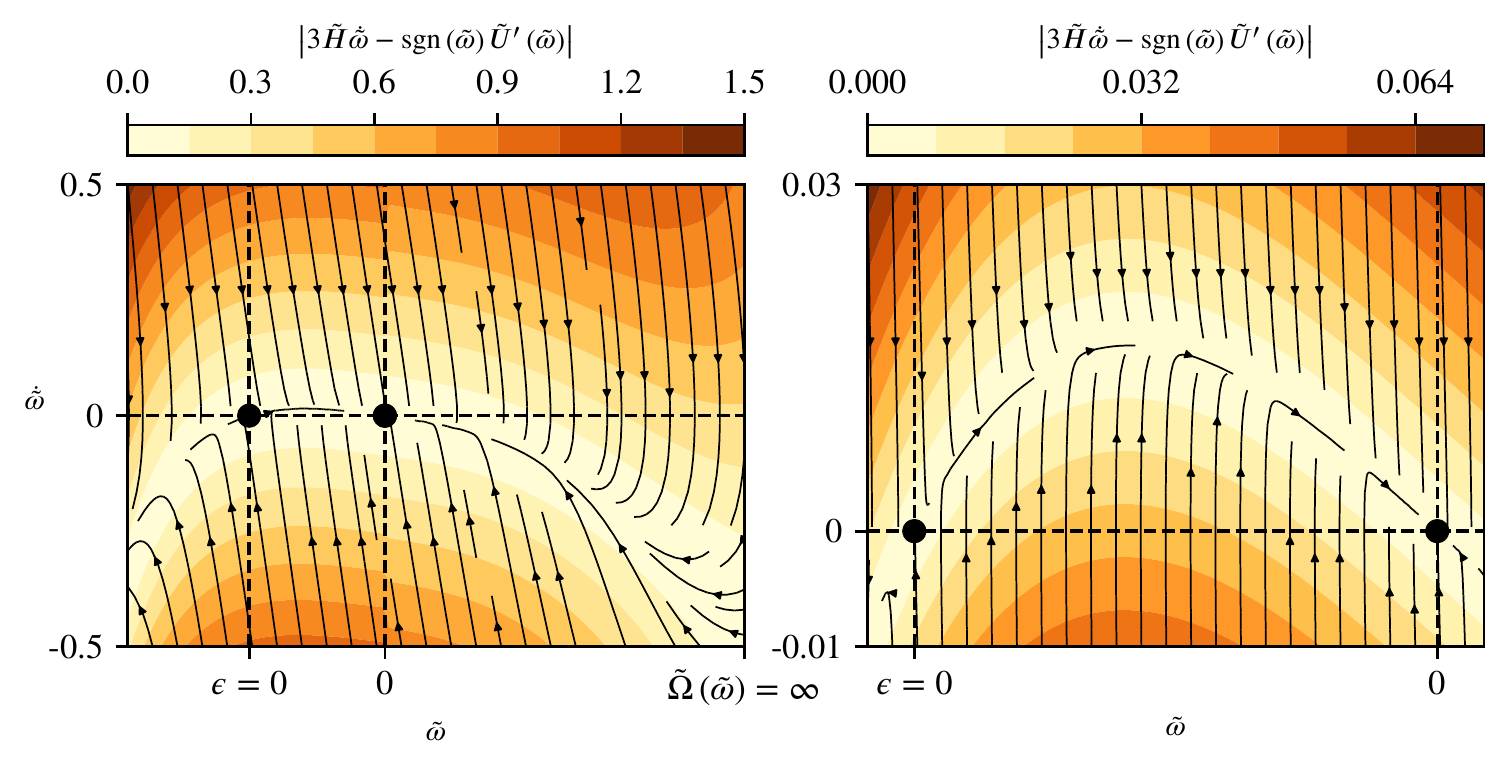}
	\caption{Phase space portrait of the inflationary potential $\efU{}$, with the attractor solutions (satisfying slow roll such that $z=0$, with $z$ defined in \cref{z SR condition}) shown by the orange dotted line and the blue dot representing the CS point, and $\Lambda \approx -0.75$, $\lambda \approx 1.75$. The colour plot shows the values of the deviation from the slow roll condition $z$.}
	\label{fig:inflationary phase space}
\end{figure*}

Following from the dynamical systems phase portraits, we can also construct a phase space with inflationary views in mind. Using the method outlined in \cite{inflation_phase_space}, we can construct a phase space that eliminated the Hubble number $H$ from the system. Using the Friedmann equations, and the Klein Gordon equation for the Lagrangian in \cref{cosh and cos lag}, which read 
\begin{align}
	3\efH{}^2 = \frac{-\alpha}{2} \Dot{\efpsi{}} + \efU{}(\efpsi{}),\quad
	\ddot{\efpsi{}} + 3\efH{}^2 \Dot{\efpsi{}} -\alpha \efU{} (\efpsi{}) ^\prime =0, 
\end{align}
where the $^\prime$ superscript denotes derivative w.r.t $\efpsi{}$, and  $\alpha = \mathrm{sgn}(\efpsi{})$ . Now the phase space can be constructed for the $\left(\efpsi{} ,\Dot{\efpsi{}} \right)$ phase space, such that 
\begin{align}
	\mathbf{x} \equiv \left(\efpsi{} ,\Dot{\efpsi{}} \right) &= \left(\Dot{\efpsi{}}, -3\efH{}^2 \Dot{\efpsi{}} +\alpha \efU{} (\efpsi{}) ^\prime\right) \nonumber\\
	& = \left(\Dot{\efpsi{}},\alpha \efU{} (\efpsi{}) ^\prime -\Dot{\efpsi{}} \sqrt{3} \sqrt{\efU{} (\efpsi{}) -\alpha \frac{\Dot{\efpsi{}}^2}{2}}\right),
\end{align}
and the full phase space, upon substituting the external cosmological constant with $\Lambda = 1-\lambda$ (where we have set $\LambdaLCDM{} =1$), is of the form 
\begin{widetext}
	\begin{align}
		\mathbf{\Dot{x}} = 
		\begin{cases}
			\left(\Dot{\efpsi{}},\frac{(\lambda -1) \sinh \left(\sqrt{2} {\efpsi{}}\right)}{4\sqrt{2}}-\sqrt{\frac{3}{8}} \Dot{\efpsi{}} \sqrt{\lambda -(\lambda -1) \cosh \left(\sqrt{2} {\efpsi{}}\right)+4 \cosh \left(\frac{{\efpsi{}}}{\sqrt{2}}\right)+4 \Dot{\efpsi{}}^2+3}-\frac{\sinh \left(\frac{{\efpsi{}}}{\sqrt{2}}\right)}{\sqrt{8}}\right), & {\efpsi{}} \leq 0,\\
			\left(\Dot{\efpsi{}},\frac{(\lambda -1) \sin \left(\sqrt{2} {\efpsi{}}\right)}{4\sqrt{2}}-\sqrt{\frac{3}{8}} \Dot{\efpsi{}} \sqrt{\lambda -(\lambda -1) \cos \left(\sqrt{2} {\efpsi{}}\right)+4 \cos \left(\frac{{\efpsi{}}}{\sqrt{2}}\right)-4 \Dot{\efpsi{}}^2+3}-\frac{\sin \left(\frac{{\efpsi{}}}{\sqrt{2}}\right)}{\sqrt{8}}\right), & 0< {\efpsi{}}< \frac{\sqrt{8} \pi}{3} ,
		\end{cases}
	\end{align}
\end{widetext}
The phase space is plotted from the hill-top at $\epsilon =0$, up to $\efpsi{} = \sqrt{8} \pi/3$ which corresponds to the upper limit of the field redefinition of $\efpsi{}$. 

From \cref{fig:inflationary phase space}, the colour function tracks the value of $\left| 3 \efH{}^2 \efpsi{} -\mathrm{sgn} (\efpsi{}) {\efU{}}^\prime (\efpsi{})  \right|$; this being the deviation from slow roll of the system. As can be seen from the plot, the phase space effectively picks out the inflationary attractor, as the system progresses towards the CS point at the origin. 

\section{Conclusions}\label{Conclusions}

In this work we demonstrated that the entire cosmology of the CTEG theory, which we defined in~\eqref{CTEG}, may be encoded in a single potential function. The CTEG is a non-Riemannian theory with curvature and torsion, which contains no Einstein--Hilbert term and therefore has no a priori connection to the IR limit of GR~\cite{chapter2,Barker_2020,mythesis}. The CTEG is independently motivated by a unitary, power-counting renormalisable particle spectrum~\cite{Lin2,Lin1}.

\subsubsection{Summary of the model}

Our main result is as follows. The CTEG is formulated in~\eqref{CTEG} with conventional minimally-coupled matter, and a metric $\tensor{g}{_{\mu\nu}}$ which is FLRW on homogeneous, isotropic scales according to~\eqref{metric}. This is consistent with the LCDM model in which the cosmological constant $\Lambda+\lambda$ is the sum of external $\Lambda$ and emergent $\lambda$ components, which are both couplings in~\eqref{CTEG}. On these scales the torsion tensor contains only the scalars $\phi$ and $\psi$ in~\eqref{TsamparlisFormula}, so that one may view the model as a torsion-free (i.e. Riemannian) scalar-tensor theory. Because $\phi$ drops out entirely as an algebraically determined quantity in the CTEG field equations, we can focus on a conformal transformation of the physical metric to $\tensor{g}{_{\mu\nu}}\equiv\efOmega{}(\efpsi{})^2\gef{_{\mu\nu}}$, where $\psi\equiv\psi(\efpsi{})$ is defined in terms of a dimensionless reparameterised field $\efpsi{}$. This is the metric of an emphatically \emph{non-physical} but nonetheless \emph{convenient} conformal frame, in which the Riemann curvature tensor is $\efR{^\mu_{\nu\sigma\lambda}}$. In the space of such conformal shifts and reparameterisations, there exists a remarkably simple picture of~\eqref{CTEG};
\begin{widetext}
	\begin{equation}\label{SummaryOfModel}
		\efL{CTEG}=
		\begin{cases}
			-\frac{\Planck^2}{2}\efR{}+\efX{\efpsi{}\efpsi{}}-\efU{}(\efpsi{})+\efL{M}
			\vphantom{\left(\frac{\efpsi{}}{\sqrt{8}}\right)},\\
			\quad\\
			-\frac{\Planck^2}{2}\efR{}-\efX{\efpsi{}\efpsi{}}-\efU{}(\efpsi{})+\efL{M}
			\vphantom{\left(\frac{\efpsi{}}{\sqrt{8}}\right)},
		\end{cases}
		\hspace{-5pt}
		\efU{}(\efpsi{})=
		\begin{cases}
			\frac{\Planck^2}{2} 
			\cosh^2\left(\frac{\efpsi{}}{\sqrt{8}}\right) \left[2 \lambda +\Lambda +\Lambda  
			\cosh \left(\frac{\efpsi{}}{\sqrt{2}}\right)\right], &\efpsi{}\leq 0,\\
			\quad\\
			\frac{\Planck^2}{2} 
			\cos^2\left(\frac{\efpsi{}}{\sqrt{8}}\right) \left[2 \lambda +\Lambda +\Lambda  
			\cos \left(\frac{\efpsi{}}{\sqrt{2}}\right)\right], & 0<\efpsi{}<\frac{\sqrt{8}\pi}{3},
		\end{cases}
	\end{equation}
\end{widetext}
where $\efX{\efpsi{}\efpsi{}}\equiv\frac{\Planck^2}{2}\efg{^{\mu\nu}}\tensor{\partial}{_\mu}\efpsi{}\tensor{\partial}{_\nu}\efpsi{}$ is a canonical kinetic term which becomes ghostly in the $0<\efpsi{}<\sqrt{8}\pi/3$ regime, and $\efU{}(\efpsi{})$ is a potential which encodes all the phenomenology. The possibility of this ghost should not be over-interpreted: the validity of the model is confirmed only at the level of the cosmological background, and so there can be no meaningful notion of particle production. The inflection and resulting concavity of the potential can then ensure that the background evolution is qualitatively similar on both sides of the origin. Due to the change in conformal frame, note that the matter Lagrangian $\efL{M}$ will inevitably acquire some dependence on $\efpsi{}$. Accordingly, it is most meaningful to use~\eqref{SummaryOfModel} in scenarios where the matter content of the Universe may be neglected. In this work we have focussed on two such regimes: hilltop inflation as the new scalar rolls slowly through some range of $\efpsi{}<0$ in the early Universe, and finally the emergence of the late asymptotic de Sitter Universe as $\efpsi{}\to 0$ from below.

How does the model~\eqref{SummaryOfModel} map to the CTEG in~\eqref{CTEG}, and why consider these values of $\efpsi{}$? The conformal shift and field redefinition required to reach the theory in~\cref{SummaryOfModel} are respectively
\begin{widetext}
	\begin{equation}\label{TransformationToModel}
		\efOmega{}(\efpsi{})=
		\begin{cases}
			\frac{9}{2} \left[\cosh \left(\frac{\efpsi{}}{\sqrt{2}}\right)+1\right]\left[2 \cosh \left(\frac{\efpsi{}}{\sqrt{2}}\right)+1\right]^{-2},\\
			\quad \\
			\frac{9}{2} \left[\cos \left(\frac{\efpsi{}}{\sqrt{2}}\right)+1\right]\left[2 \cos \left(\frac{\efpsi{}}{\sqrt{2}}\right)+1\right]^{-2} ,
		\end{cases}
		\quad\quad
		\left|\psi(\efpsi{})\right|=
		\begin{cases}
			\psi_{\text{C}}\mathrm{sech}\left(\frac{\efpsi{}}{\sqrt{8}}\right), &\efpsi{}\leq 0,\\
			\quad\\
			\psi_{\text{C}}\mathrm{sec}\left(\frac{\efpsi{}}{\sqrt{8}}\right), & 0<\efpsi{}<\frac{\sqrt{8}\pi}{3}.
		\end{cases}
	\end{equation}
\end{widetext}
The modulus is expected in~\eqref{TransformationToModel} because the CTEG cosmological field equations are not sensitive to $\mathrm{sgn}(\psi)$.
From the transformation in~\eqref{TransformationToModel}, we identify the negative semidefinite range ${\efpsi{}\in\left(-\infty,0\right]}$ monotonically with ${\psi\in\left(0,\psi_{\text{C}}\right]}$, wherein the background axial torsion magnitude grows from the neighbourhood of zero up to its critical condensate value $\psi_{\text{C}}$ defined in~\eqref{TorsionCondensate}. Following~\cite{chapter2} and as discussed in~\cref{TheCTEGModel}, we speculate that there is an extended pause at some $\psi\lesssim\psi_{\text{C}}$ during the radiation-dominated epoch\footnote{Note an \emph{erroneous} remark in~\cite{Barker_2020} claiming the pause is at $\psi\gtrsim\psi_{\text{C}}$.}. Since the final approach to $\psi\to\psi_{\text{C}}$ is known to be over-damped by the Hubble friction~\cite{chapter2,Barker_2020,mythesis}, it is reasonable to assume that the Universe inhabits this range throughout its whole history\footnote{The range ${\efpsi{}\in\left(0,\sqrt{8}\pi/3\right)}$ spans increasing values of the torsion ${\psi\in\left(\psi_{\text{C}},2\psi_{\text{C}}\right)}$ between one and two times the level of the condensate. The conformal shift ranges from ${\efOmega{}(\efpsi{})\in\left(1,\infty\right)}$ and is not defined beyond this point.}. In this range, the scalar $\efpsi{}$ appears with a positive kinetic energy in~\eqref{SummaryOfModel}: it is a conventional scalar whose phenomenology is wholly determined by its potential $\efU{}(\efpsi{})$. As $\psi$ increases from zero towards $\psi_{\text{C}}$, so the conformal shift increases through the range $\efOmega{}(\efpsi{})\in\left(0,1\right]$. It is suggestive that the conformal frames are \emph{equivalent} at the condensate, where $\efpsi{}$ vanishes.

\subsubsection{Strong coupling from dark energy?}

The particle spectrum of CTEG was initially computed without matter, or any external cosmological constant $\Lambda=\efL{M}=0$~\cite{Lin2,Lin1}. For torsion below the level of the condensate, the potential then takes the very natural form $\efU{}(\efpsi{})=\Planck^2\lambda\cosh^2\left(\efpsi{}/\sqrt{8}\right)$. This empty Universe inevitably evolves towards the condensate ${\psi\to\psi_{\text{C}}}$ at the bottom of the potential ${\efpsi{}\to 0}$, which drives an asymptotically de Sitter expansion in the asymptotically coincident conformal frames ${\efH{}\to H\to\sqrt{\lambda/3}}$. This is perfectly consistent with the previous nonlinear analyses~\cite{Barker_2020,mythesis}. However, we recall that the motivating particle spectrum analysis assumed a background value of \emph{zero} torsion. The limit ${\psi\to 0}$ corresponds to the limit ${\efpsi{}\to -\infty}$, which sends us to the `top' of the hyperbolic cosine potential in~\cref{fig:inflation plot}.

This is an unsettling result. Around the zero-torsion Minkowski background, the first-order perturbations to the $\text{PGT}^{q+}$ Lagrangian vanish (or are pure surface terms), confirming this background to be a solution of the nonlinear field equations. The second-order perturbations give rise to a unitary, power-counting renormalisable particle spectrum~\cite{Lin2}. It is not however clear what perturbative treatment can be valid near the divergent potential at negative infinity. The existence of this potential was only revealed in the current work, using inherently non-perturbative methods which study the bulk nonlinear phase space of the theory.

A possible interpretation is that the zero-torsion Minkowski background is a \emph{strongly coupled} surface of the theory~\cite{BeltranJimenez:2020lee,smooth}. Many $\text{PGT}^{q+}$ Lagrangia are well known to be spoiled by the strong coupling effect~\cite{2002IJMPD..11..747Y,chapter4,mythesis}. Near such surfaces, perturbative methods cannot apply, and they yield a particle spectrum which belongs to a `fictitious' theory. The effect is especially dangerous because no indication of strong coupling can arise at the perturbative order used in the propagator analysis: the method `fails silently'.

A thorough investigation into the current case will take into account not only the divergent potential, but also the vanishing conformal factor~\eqref{TransformationToModel} in the limit $\efpsi{}\to -\infty$. For the moment we note from~\eqref{TransformationToModel} that, in the absence of other non-minimally coupled matter, a non-divergent potential strictly requires $\lambda=\Lambda=0$. In this case $\efpsi{}$ becomes shift-symmetric, and no shadow is cast by the current work on the validity of the zero-torsion particle spectrum: as a phenomenological consequence all forms of dark energy are, however, forfeit.

In this context it may be appropriate to relax the interpretation of $\lambda$ and $\Lambda$ as bare couplings in the theory. These might run, or be anomalously acquired in an effective field theory framework, in a way which should be shown to be consistent with the perturbative QFT. We leave this investigation to future work.

\subsubsection{Hilltop inflation}

Dynamically adjusted (or renormalised) $\lambda$ and $\Lambda$ may also be necessary for inflation. A key result in the current treatment, which resolves some speculation in previous work~\cite{chapter2,Barker_2020,mythesis}, is that a concrete inflationary model emerges entirely within the gravitational sector of CTEG. As illustrated in~\cref{fig:inflation plot}, if $\Lambda<0$ and $\lambda>0$ then $\efU{}(\efpsi{})$ can become a hilltop potential for $\efpsi{}<0$. 

As part of this work, we demonstrated in~\cref{fig:phase space JF with critical points,fig:EF phase space}, using various conformal frames, that the torsion condensate is a late-Universe attractor provided $-2 \nleq \lambda/\Lambda \nleq -1$.
In this case the asymptotic de Sitter expansion in the late Universe is ${\efH{}\to H\to\sqrt{(\lambda+\Lambda)/3}}$. In principle, values $\lambda = 4.987\times 10^{-122}\Planck^2$ and $\Lambda =  -2.846\times 10^{-122}\Planck^2$ can be found which generate 50 e-folds of inflation in the hilltop slow-roll regime, whilst remaining consistent with current estimates of the accelerated expansion at late times~\cite{2018arXiv180706209P}. Yet these parameters are not legitimate for the following reason. As can be seen from~\cref{fig:inflation plot}, the scale of inflation with these parameters is inseparable from that of the current Hubble number: the early- and late-Universe phenomena cannot be `resolved'.

For the moment therefore, the slow-roll dynamics demonstrated in~\cref{fig:inflationary phase space} are encouraging, but should be viewed \emph{qualitatively}. Equipped with the new reduced theory in~\cref{SummaryOfModel,TransformationToModel}, our understanding of the classical background phenomena of CTEG in~\cref{CTEG} is unlikely now to be improved by further study. Progress instead rests on the interpretation of CTEG as an effective quantum theory in which torsion and matter are coupled.

\begin{pquotation}{To The Lighthouse, Virginia Woolf, 1927}
	``The great revelation perhaps never did come. Instead there were little daily miracles, matches struck unexpectedly in the dark; here was one.''
\end{pquotation}

\begin{acknowledgements} 
	
	We are grateful for insightful discussions with Anthony Lasenby, Mike Hobson and Will Handley.
	
	C.R is grateful for the opportunity of the summer internship with the Institute of Astronomy (IoA) which facilitated this work. W.E.V.B. is grateful for the kind hospitality of Leiden University and the Lorentz Institute, and the support of Girton College, Cambridge.
	
\end{acknowledgements}

	\bibliography{Manuscript}
	
\end{document}